\input harvmac
%
%
\input epsf.tex


%
%
\ifx\epsfbox\UnDeFiNeD\message{(NO epsf.tex, FIGURES WILL BE IGNORED)}
\def\figin#1{\vskip2in}
\else\message{(FIGURES WILL BE INCLUDED)}\def\figin#1{#1}
\fi
\def\Fig#1{Fig.~\the\figno\xdef#1{Fig.~\the\figno}\global\advance\figno
 by1}
%
%
%
%
\def\ifig#1#2#3#4{
\goodbreak\midinsert
\figin{\centerline{\epsfysize=#4truein\epsfbox{#3}}}
\narrower\narrower\noindent{\footnotefont
{\bf #1:}  #2\par}
\endinsert 
}

%

%
%
\def\ajou#1&#2(#3){\ \sl#1\bf#2\rm(19#3)}
\def\jou#1&#2(#3){,\ \sl#1\bf#2\rm(19#3)}

\def\ap{{\it Ann. Phys, NY}}
\def\cqg{{\it Class. Quant. Grav.}}
\def\cmp{{\it Comm. Math. Phys.}}

\def\jmp{{\it J. Math. Phys.}}
\def\ijmp{{\it Int. J. Mod. Phys. }}

\def\np{{\it Nucl. Phys.}}

\def\pl{{\it Phys. Lett.}}
\def\pr{{\it Phys. Rev.}}
\def\prl{{\it Phys. Rev. Lett.}}

\def\atmp{{\it Adv. Theor. Math. Phys.}}

%
%
\font\bbbi=msbm10
\font\ticp=cmcsc10
%
%

\def\eg{{\it e.g.}}
\def\p{\partial}

\def\eg{{\it e.g.}}
\def\cf{{\it c.f.}}

\def\apr{\alpha'}

\def\half{\frac{1}{2}}
\def\bbb#1{\hbox{\bbbi #1}}
\def\vecm{{\vec m}}
\def\gbdel{{G_{B\partial}}}
\def\calo{{\cal O}}

\def\rst{{r_*}}
\def\bra#1{{\langle #1 |}}
\def\ket#1{{| #1 \rangle}}

\def\vol2{d^{2}\sigma} 
\def\frac#1#2{{#1 \over #2}}


\def\gym{g_{\rm YM}}
\def\ads#1{{\rm AdS}_{#1}}
\def\unlockat{\catcode`\@=11}
\def\lockat{\catcode`\@=12}
\def\cft#1{{\rm CFT_{#1}}}

\unlockat

\def\newsec#1{\global\advance\secno
by1\message{(\the\secno. #1)}
\global\subsecno=0\global\subsubsecno=0\eqnres@t\noindent
{\bf\the\secno. #1}
\writetoca{{\secsym} {#1}}\par\nobreak\medskip\nobreak}
%
\global\newcount\subsecno \global\subsecno=0
\def\subsec#1{\global\advance\subsecno
by1\message{(\secsym\the\subsecno. #1)}
\ifnum\lastpenalty>9000\else\bigbreak\fi\global\subsubsecno=0
\noindent{\bf \secsym\the\subsecno. #1}
\writetoca{\string\quad {\secsym\the\subsecno.} {#1}}
\par\nobreak\medskip\nobreak}
%

\global\newcount\subsubsecno \global\subsubsecno=0
\def\subsubsec#1{\global\advance\subsubsecno by1
\message{(\secsym\the\subsecno.\the\subsubsecno. #1)}
\ifnum\lastpenalty>9000\else\bigbreak\fi
\noindent{\bf \quad{\secsym\the\subsecno.\the\subsubsecno.}\ {#1}}
\writetoca{\string\qquad{\secsym\the\subsecno.\the\subsubsecno.}{#1}}
\par\nobreak\medskip\nobreak}


\def\subsubseclab#1{\DefWarn#1\xdef
#1{\noexpand\hyperref{}{subsubsection}%
{\secsym\the\subsecno.\the\subsubsecno}%
{\secsym\the\subsecno.\the\subsubsecno}}%
\writedef{#1\leftbracket#1}\wrlabeL{#1=#1}}
\lockat


\def\paragraph#1{\nobreak\medskip\nobreak\noindent {\bf #1}}

\hyphenation{Min-kow-ski}
%
%
%
%
%
\lref\steveaki{S.S. Gubser and A. Hashimoto, ``Exact absorption
probabilities for the D3-brane'', hep-th/9805140.}

\lref\gkp{S.S. Gubser, I. Klebanov and A. Polyakov, ``Gauge
theory correlators from noncritical string theory'', hep-th/9802109,
\pl\ {\bf B428} (1998) 105.}

\lref\Maldconj{J. Maldacena, ``The large-N limit of superconformal
filed theories and supergravity'', hep-th/9711200, \atmp\ {\bf 2}
(1998) 231.}

\lref\Wittads{E. Witten, ``Anti-de Sitter space and holography'',
hep-th/9802150, \atmp\ {\bf 2} (1998) 253.}

\lref\BKL{V. Balasubramanian, P. Kraus and A. Lawrence,
``Bulk vs. Boundary dynamics in anti-de Sitter spacetime'',
hep-th/9805171, \pr\ {\bf D59:046003} (1999).}

\lref\cilar{S.B. Giddings, ``Comments on information loss and remnants,''
hep-th/9310101, \pr\ {\bf D49} (1994) 4078.}

\lref\BKLT{V. Balasubramanian, P. Kraus, A. Lawrence and S. Trivedi,
``Holographic probes of anti-de Sitter space-times'',
hep-th/9808017, Harvard preprint HUTP-98/A057, Caltech preprint
CALT68-2189, Fermilab preprint Pub-98/240-T}

\lref\BDHM{T. Banks, M. Douglas, G. Horowitz and E. Martinec,
``AdS dynamics from conformal field theory'', hep-th/9808016,
UCSB/ITP preprint NSF-ITP-98082, U. Chicago preprint EFI-98-30.} 

\lref\FMMR{D.Z. Freedman, S.D. Mathur, A. Matusis and L. Rastelli,
``Correlation functions in the $\cft{d}/\ads{d+1}$ correspondence'',
hep-th/9804058, MIT preprint MIT-CTP-2727.}

\lref\joesmatrix{J. Polchinski, ``S-matrices from AdS spacetime'',
hep-th/9901076.}

\lref\lennysmatrix{L. Susskind, ``Holography in the flat-space 
limit'', hep-th/9901079.}

\lref\batemantwo{A. Erd\'elyi, ed., {\it Higher Transcendental
Functions} vol. 2, Kreiger (Malabar, Florida) 1981.}

\lref\breitfreed{P. Breitenlohner and D.Z. Freedman,
``Positive energy in anti-de Sitter backgrounds and gauged
extended supergravity'', \pl\ {\bf B115} (1982) 197;
``Stability in gauged extended supergravity'', \ap\ {\bf 160} 
(1985) 406.}

\lref\meztown{L. Mezinescu and P.K. Townsend, ``Stability at a local
maximum in higher-dimensional anti-de Sitter space and applications to
supergravity", \ap\ {\bf 160} (1985) 406.}

\lref\susswitt{L. Susskind and E. Witten, ``The holographic
bound in anti-de Sitter space'', hep-th/9805114, Stanford preprint
SU-ITP-98-39.}

\lref\amandajoe{A. Peet and J. Polchinski, ``UV/IR relations
in AdS dynamics'', hep-th/9809022, UCSB/ITP preprint NSF-ITP-98-086}

\lref\romansetal{H.J. Kim, L. Romans and P. 
van Nieuwenhuizen, ``Mass spectrum of chiral ten-dimensional N=2
supergravity on $S^5$'', \pr\ {\bf D32} (1985) 389.}

\lref\weinbergone{S. Weinberg, {\it The Quantum
Theory of Fields}, vol. 1, Cambridge University Press (1995).}

\lref\banksgreen{T. Banks and M. Green, ``Non-perturbative
effects in $\ads{5}\times S^5$ and $d=4$ SUSY Yang-Mills'',
hep-th/9804170, {\bf JHEP 9805:002} (1998).}

\lref\hawkingellis{S. Hawking and G.F.R. Ellis, {\it The large-scale
structure of spacetime}, Cambridge University Press (1973).}

\lref\luschermack{M. L\"uscher and G. Mack,
``Global conformal invariance in quantum field theory'',
\cmp\ {\bf 41} (1975) 203.}

\lref\holog{G. 't Hooft, ``Dimensional reduction in
quantum gravity'', gr-qc/9310026, Salamfest 1993:0284-296;
L. Susskind, ``The world as a hologram'', hep-th/9409089,
\jmp\ {\bf 36} (1995) 6377.}

\lref\greybody{S.S.~Gubser, I.R.~Klebanov and A.A.~Tseytlin, 
``String theory and classical absorption by three-branes", 
\np\ {\bf B499} (1997) 217, hep-th/9703040;
M.~Cvetic and F.~Larsen, ``Grey body factors for rotating 
black holes in four-dimensions", \np {\bf B506} (1997) 107, 
hep-th/9706071; S.D.~Mathur and A.~Matusis, 
``Absorption of partial waves by three-branes", hep-th/9805064.}

\lref\klebrev{I.R. Klebanov, 
``From threebranes to large N gauge theories", hep-th/9901018,
Princeton preprint PUPT-1831.}

\lref\klebnon{I.R. Klebanov, ``World volume approach to absorption by
non-dilatonic branes", hep-th/9702076, \np\ {\bf B496} (1997) 231.} 

\lref\iandz{C. Itzykson and J.-B. Zuber, {\it Quantum Field Theory},
McGraw-Hill (NY) 1980.}

\lref\capelli{A. Capelli, D. Friedan and J.I. Latorre,
``c-theorem and Spectral Representation'', \np\ {\bf B352}
(1991) 616.}

\lref\dusedau{D.W. D\"usedau and D.Z. Freedman,
``Lehmann spectral representation for anti-de
Sitter quantum field theory'', \pr\ {\bf D33} (1986) 389.}

\lref\absteg{M. Abramowitz and I.A. Stegun,
{\it Handbook of Mathematical Functions}, Dover (NY) 1965.}

\lref\garysimon{G. Horowitz and S. Ross, ``Possible resolution
of black hole singularities from gauge theory'', hep-th/9803085,
JHEP 9804:15 (1988)}

\lref\juanandy{J. Maldacena and A. Strominger,
``$\ads{3}$ black holes and a stringy exclusion principle'',
hep-th/9804085, Harvard preprint HUTP-98/A016.}

\lref\avisetal{S.J. Avis, C.J. Isham and D. Storey,
``Quantum field theory in anti-de Sitter space-time'',
\pr\ {\bf D18} (1978) 3565.}

\lref\gregandnati{G. Moore and N. Seiberg, ``From
loops to fields in 2D quantum gravity'', \ijmp\ {\bf A7}
(1992) 2601.}

\lref\oferanded{O. Aharony and E. Witten, ``Anti-de
Sitter space and the center of the gauge group'',
hep-th/9807205, {\bf JHEP 9811:018}\ (1998).}

\lref\ght{G.W. Gibbons, G.T. Horowitz and
P.K. Townsend, ``Higher-dimensional
resolution of dilatonic black hole singularities'',
hep-th/9410073, \cqg {\bf 12} (1995) 297.}

\lref\emilmat{E.J. Martinec, ``Matrix models
of AdS gravity'', hep-th/9804111, U. Chicago
preprint EFI-98-13.}

\lref\jacobson{T. Jacobson, ``Thermodynamics of space-time:
the Einstein equation of state'', gr-qc/9504004,
\prl\ {\bf 75} (1995) 1260.}

\lref\emilstat{E.J. Martinec, ``Conformal field
theory, geometry, and entropy'', hep-th/9809021,
U. Chicago preprint EFI-98-40.}

\lref\jevicki{A. Jevicki and B. Sakita, ``The quantum
collective field method and its application to the
planar limit'', \np\ {\bf B165} (1980) 511.}

\lref\oferetal{O. Aharony, M. Berkooz, D. Kutasov and N. Seiberg,
``Linear dilatons, NS5-branes and holography",
hep-th/9808149, {\bf JHEP 9810:004} (1998)}

\baselineskip=15pt plus 2pt minus 2pt 
\Title{\vbox{\baselineskip-12pt\hbox{HUTP-99/A005, NSFITP-99-06}
\hbox{hep-th/9902052}
}} 
{\vbox{\centerline {What Do CFTs Tell Us About}
\medskip
\centerline{Anti-de Sitter Spacetimes?}}}
\centerline{\ticp Vijay Balasubramanian,${}^{1,2}$\footnote{$^*$}
{vijayb@pauli.harvard.edu.  Permanent address: Harvard University} 
Steven
B. Giddings,${}^{3}$\footnote{$^\dagger$}{giddings@physics.ucsb.edu} 
and Albion
Lawrence${}^{1,4}$\footnote{$^\ddagger$}{lawrence@cartan.harvard.edu.
Permanent address: Harvard University.}}
\smallskip
\centerline{${}^1$\sl Lyman Laboratory of Physics, Harvard University}
\centerline{\sl Cambridge, MA 02138}
\smallskip
\centerline{${}^2$\sl Institute for Theoretical 
Physics, University of California}
\centerline{\sl Santa Barbara, CA 93106}
\smallskip
\centerline{${}^3$ \sl Department of Physics, University of California,}
\centerline{\sl Santa Barbara, CA 93106}
\smallskip
\centerline{${}^4$ \sl Department of Physics and Astronomy, 
Rutgers University,}
\centerline{\sl Piscataway, NJ 08855}
\bigskip
\centerline{\bf Abstract}
The AdS/CFT conjecture relates quantum gravity on Anti-de Sitter (AdS)
space to a conformal field theory (CFT) defined on the spacetime
boundary.  We interpret the CFT in terms of natural analogues of the
bulk S-matrix.  Our first approach finds the bulk S-matrix as a limit
of scattering from an AdS bubble immersed in a space admitting
asymptotic states.  Next, we show how the periodicity of geodesics
obstructs a standard LSZ prescription for scattering within global
AdS.  To avoid this subtlety we partition global AdS into patches
within which CFT correlators reconstruct transition amplitudes of AdS
states.  Finally, we use the AdS/CFT duality to propose a large $N$
collective field theory that describes local, perturbative
supergravity.  Failure of locality in quantum gravity should be
related to the difference between the collective $1/N$ expansion and
genuine finite $N$ dynamics.
\Date{}

\newsec{Introduction}

The AdS/CFT correspondence states that string theory in Anti-de Sitter
(AdS) spacetime is ``holographically" dual to a conformal field theory
(CFT) defined on the spacetime boundary \refs{\Maldconj\gkp-\Wittads}.
The semiclassical limit of spacetime physics is related to the large N
limit of the CFT.  Semiclassical spacetime physics is generally
expected to satisfy the familiar axioms of locality and causality.  On
the other hand, some sort of violation of locality is expected for a
holographic quantum gravity that satisfies the Bekenstein bound
\holog.  For example, anti-de Sitter spacetime locality appears to
translate into scale locality in the dual CFT, a property that is
certainly absent for finite N \refs{\susswitt,\BDHM,\BKLT}.  Given
holography, we are therefore faced, nearly inevitably, with violations
of locality in semiclassical quantum gravity.

To investigate the failure of locality, we must
understand what data about spacetime physics are accessible from
CFT calculations.  As summarized in Sec.~2, the AdS/CFT
correspondence equates the Anti de-Sitter effective action, seen as a
functional of boundary data, to the generating functional for
correlators of a CFT defined on the spacetime boundary
\refs{\gkp,\Wittads}.  All  information  available in the  CFT is
contained in these correlation functions, evaluated in all the
possible states of the theory.  We will show that these data
reconstruct bulk transition amplitudes.

In AdS spacetimes of lorentzian signature, solutions to the free bulk
wave equations can be classified into ``normalizable'' and
``non-normalizable'' modes encoding, respectively, the states of the
theory and the boundary conditions for fields \BKL.  A conventional
LSZ prescription for spacetime physics would describe transition
amplitudes between asymptotic states, in terms of truncated Green
functions integrated against normalizable modes.  AdS lore states that
asymptotic states cannot be defined, due to the timelike boundary and
the periodicity of geodesics.  However, in Sec.~3 we will show that
the vacuum correlation functions of the dual CFT can be given an
S-matrix interpretation as a limit of scattering from an AdS
``bubble'' immersed in a space admitting asymptotic states.  In
Poincar\'e coordinates for $\ads{5}$, this is simply the usual CFT
interpretation of scattering from an asymptotically flat 3-brane
\refs{\klebnon,\klebrev}.  Global AdS does not arise this way 
as a limit of a conventional brane solution.  Therefore, for
interpretational purposes, we construct a global AdS bubble inside a
simple asymptotic spacetime.  A stable string solution of this
kind is not known. Nevertheless, by studying scattering in this
spacetime, we will argue that the CFT correlators recover universal
features of an S-matrix for scattering in global AdS.  Interestingly,
our interpretation identifies certain singularities of the CFT
correlators with resonant scattering from the Anti de-Sitter region.

The results of Sec.~3 arise because CFT correlators are expressed in
the bulk as truncated Green functions convolved against
non-normalizable modes.  In Sec.~4 we return to the issue of
normalizable modes, or states localized {\it within} AdS space, and
transitions between them.  We start by showing explicitly that the
periodicity of geodesics in AdS obstructs a conventional definition of
asymptotic states. Thus the LSZ prescription is ill-defined.  Instead,
we partition global AdS into Poincar\'e patches within which geodesics
do not reconverge.  The AdS boundary is likewise partitioned, and
physics within a given bulk patch is dual to a CFT defined on the
corresponding boundary.  The in and out states of the theory on the
patch correspond to boundary conditions at early and late times.  We
show that transition amplitudes between these states are described by
correlation functions of the dual CFT.  In the large radius limit for
AdS, this construction provides a holographic description of a flat
space S-matrix. (While this paper was being written we received
\refs{\joesmatrix,\lennysmatrix} which contain a related derivation.)
The intermediate steps of our discussion rely on the diagrammatic
expansion of the spacetime physics, but we expect that the final
results are defined in the full theory.

In Sec.~5 we attempt to reconstruct local bulk operators from
the CFT.  After reviewing the $N\to\infty$ case where this
reconstruction is manifest \refs{\BDHM,\BKLT}, we discuss the problem
at finite $N$.  Using a Lehmann representation described by D\"usedau
and Freedman \dusedau, we show that finite $N$ bulk operators have a
complicated ``multi-particle'' structure and that the CFT primaries
only capture the ``single-particle'' piece.  To reconstruct 
supergravity, we propose a ``collective field theory'' built from the
spectral decomposition of large-$N$ conformal primaries in the CFT.
We are able to posit an effective action order by order in $1/N$ and
the string coupling which reproduces local, perturbative spacetime physics after a simple Bessel transformation of collective field correlators.
Presumably this collective field procedure reproduces the finite $N$
CFT at best in a power series.  We close with a discussion of the
problems facing reconstruction of local supergravity from the exact
finite $N$ dynamics.

\newsec{Review of the AdS/CFT correspondence}
\def\b{{\bf b}}
\def\x{{\bf x}}
\def\k{{\bf k}}

Euclidean AdS (EAdS) space is topologically a ball, with the metric:
\eqn\eads{
ds^2 = {R^2 \over z^2} (dt^2 + d\vec{x}^2 + dz^2)\ .
}
Here $\b \equiv \{t,\vec{x}\}$ spans ${\bbb R}^{d}$ and $\infty \leq z \leq
0$ with a boundary at $z=0$.  The EAdS/CFT correspondence
states~\refs{\gkp,\Wittads}:
\eqn\gkpw{
	Z_{{\rm bulk}} [\phi(\phi_0)] = \langle 
		e^{-\int_\partial d^d \b \, \phi_0 (\b) \, \CO(\b)}
	\rangle\ .
}
Here $\ln Z_{\rm bulk}$ is the effective action for string theory on
$\ads{d+1}$ considered as a functional of the boundary data $\phi_0$
for the fields $\phi$.  The right hand side is the generating
functional of correlators of the operator $\CO$ dual to $\phi$.
For illustration, we will always take $\phi$ to be a scalar field of
mass $m$. Then, setting  
\eqn\dimensions{
2h_{\pm} = {d \over 2} \pm \nu ~~~~~~~;~~~~~~~
\nu = {1\over 2} \sqrt{d^2 + 4m^2}\ ,
}
regular classical solutions of the free wave equation for $\phi$ have
the boundary behaviour 
\eqn\bbeh{
	\lim_{z\to 0} \phi(\b,z) = z^{2h_-} \phi_0(\b)\ ,
}
and couple to CFT operators ${\cal O}$ of dimension $2h_+$.    The
free classical solution may be suggestively written in terms of the
boundary value $\phi_0$ and a bulk-boundary propagator $G^E_{B\partial}$ as
\Wittads:
\eqn\bdyblk{\phi(\x) = \int_\partial d\b \, G^E_{B\partial}(\x,\b) \, \phi_0(\b)\ }
with $\x \equiv \{z,t,\vec{x}\}$. Using this expression and \gkpw , the
CFT correlators are 
given in terms of truncated bulk Green functions as:
\eqn\euccorr{\langle \calo(\b_1)\calo(\b_2)\cdots\calo(\b_n)\rangle = \int
\prod_{i=1}^n \left[dx_i \, G^E_{B\partial}(\b_i,\x_i)\right] \langle \phi(\x_1)\cdots
\phi(\x_n)\rangle_T \ .}
The EAdS diagrams introduced in \Wittads\ summarize this
computation.   In Fig.~1 a diametric slice of EAdS
is displayed, and the thick line is the boundary of the resulting
disc.  The CFT correlators are then obtained by replacing the legs of
ordinary bulk Feynman diagrams by bulk-boundary
propagators.\foot{Ref.~\refs{\BDHM} and section 2.3 show that they
can also be obtained from the limiting behavior of bulk diagrams as the
external points are taken to the boundary.}

\ifig{Fig.~1}{EAdS diagrams computing euclidean CFT
correlators}{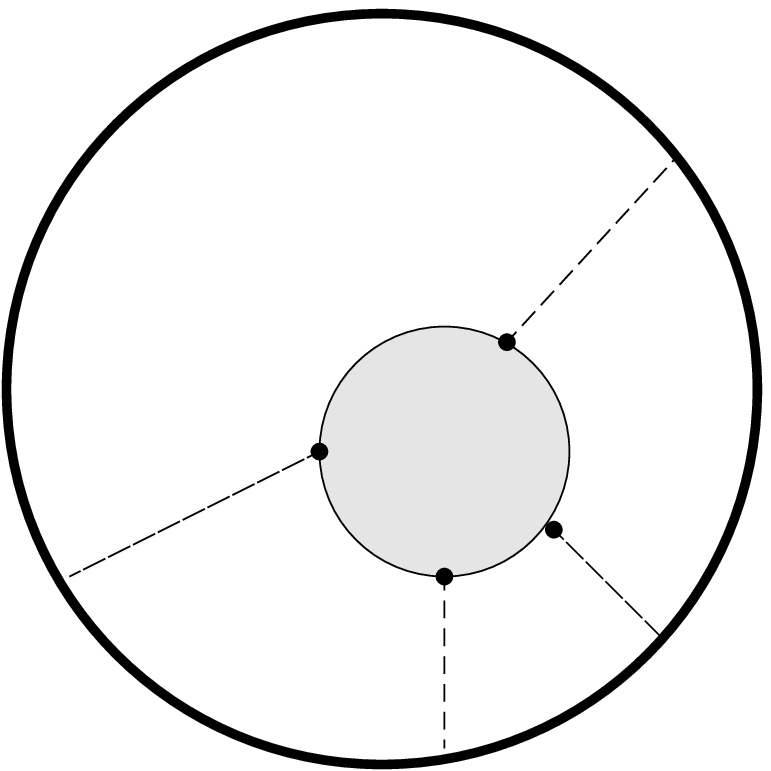}{1.0} 

\paragraph{Lorentzian correspondence:} The universal cover of lorentzian AdS
(CAdS) is topologically a cylinder, with the metric:
\eqn\cads{
ds^2 = R^2 ( -\sec^2\rho \, d\tau^2 + \sec^2\rho \, d\rho^2 +
\tan^2\rho \, d\Omega^2_{d-1} )\ .
}
The boundary of spacetime at $\rho = \pi/2$ has the topology $S^{d-1}
\times R$.    A free massive scalar field on this background 
will have regular solutions of the form:
\eqn\lorphi{
	\phi(x) = \int d\b \, \gbdel(\x,\b) \, \phi_0 (\b) + \phi_n (\x)
}
where $\phi_n$ is normalizable in the Klein-Gordon norm and vanishes
at the spacetime boundary.  The bulk-boundary propagator $\gbdel$ is a
solution to the bulk wave equation which approaches a delta function
on the AdS boundary.  This propagator is ambiguous in lorentzian AdS
since we may always add a normalizable solution $\phi_n$ to it without
changing the boundary behaviour.  We will pick the propagator arising
via continuation from euclidean AdS. This corresponds to a certain choice
of vacuum as discussed in Sec.~3.  Then \lorphi\ describes the
corresponding general
bulk solution that approaches $\phi_0$ at the boundary.

It was argued in \BKL\ that the normalizable, fluctuating solutions
$\phi_n$ encode the states of the bulk subject to fixed boundary
conditions specified 
by $\phi_0$.    The solutions $\phi_n$ provide a unitary
representation of the conformal group that matches the states we
expect the dual operator to create in the boundary CFT.   In the free
limit, {\it classical} bulk modes $\phi_n$ should be dual to
``coherent" states in the boundary.  In this limit the CAdS/CFT
correspondence is written as \BKLT: 
\eqn\gkpwclass{Z_{{\rm cl}}[\phi(\phi_0) + \phi_n] = \langle \phi_n |
e^{i\int_\partial d\b \,  \phi_0(\b)  \, \calo(\b)} | \phi_n \rangle\  
} 
where the bulk action is evaluated in the presence of the classical mode
$\phi_n$ and the CFT is placed in corresponding ``coherent'' state
$|\phi_n\rangle$.   In this limit we do not interpret the computation
diagramatically and instead (semiclassically) 
evaluate the bulk action by using the
equations of motion.  It was shown in \BKLT\ that the classical
``probe'' $\phi_n$ induces expectation values for operators in the
dual CFT.

Here we are interested in the interpretation of the normalizable modes
as asymptotic states in transition amplitudes for the bulk theory.  
To have such an interpretation we must either be able to turn off bulk
interactions at early and late times or separate the wavepackets by
large distances.  Since the effective bulk coupling constant is
determined by the constant string coupling and AdS curvature, we are
unable to change the asymptotic strength of the interaction.   But in
CAdS it is also not possible to separate wavepackets asymptotically
because of the periodicity of geodesics in spacetime.
Alternatively, the normalizable mode solutions of CAdS have fixed
temporal periodicity \refs{\breitfreed,\avisetal,\meztown,\BKLT}.  
So, if two wavepackets interact in a
near collision and and then appear to separate, they eventually bounce
off the AdS effective potential near the boundary and almost collide
again.  On the other hand, there should be states in the bulk theory
that are dual to the CFT 
states that arise from operators acting on the vacuum at early and late
times.    We will discuss these issues in detail in Sec.~4.

For the moment, we will work in the CAdS vacuum by
setting all the modes $\phi_n$ to zero.  Then the CAdS/CFT correspondence 
is given by:
\eqn\gkpwvac{Z[\phi(\phi_0)] = \langle T \, e^{i\int_\partial d\b \, \phi_0(\b)
\, \calo(\b)}\rangle\ ,}
and the CFT vacuum correlators are expressed in terms of truncated
bulk Green functions as:
\eqn\gkpwd{\langle T \, \calo(\b_1)\calo(\b_2)\cdots\calo(\b_n)\rangle = \int
\prod_{i=1}^n \left[d\x_i \, \gbdel(\b_i,\x_i)\right] \langle
T \, \phi(\x_1)\cdots 
\phi(\x_n)\rangle_T \ .}
The corresponding CAdS diagram appears in Fig.~2a.  Once again, a
diametric slice of CAdS is presented and the thick lines represent the
cylindrical AdS boundary. In Sec.~4 we will discuss how \gkpwvac\ is
modified to account for the states of AdS and transitions between
them.  Essentially, this will result in CAdS diagrams with extra legs
(Fig.~2b, for example) representing the influence of in and out
states on CFT correlators.

\ifig{Fig.~2}{CAdS diagrams computing lorentzian
correlators}{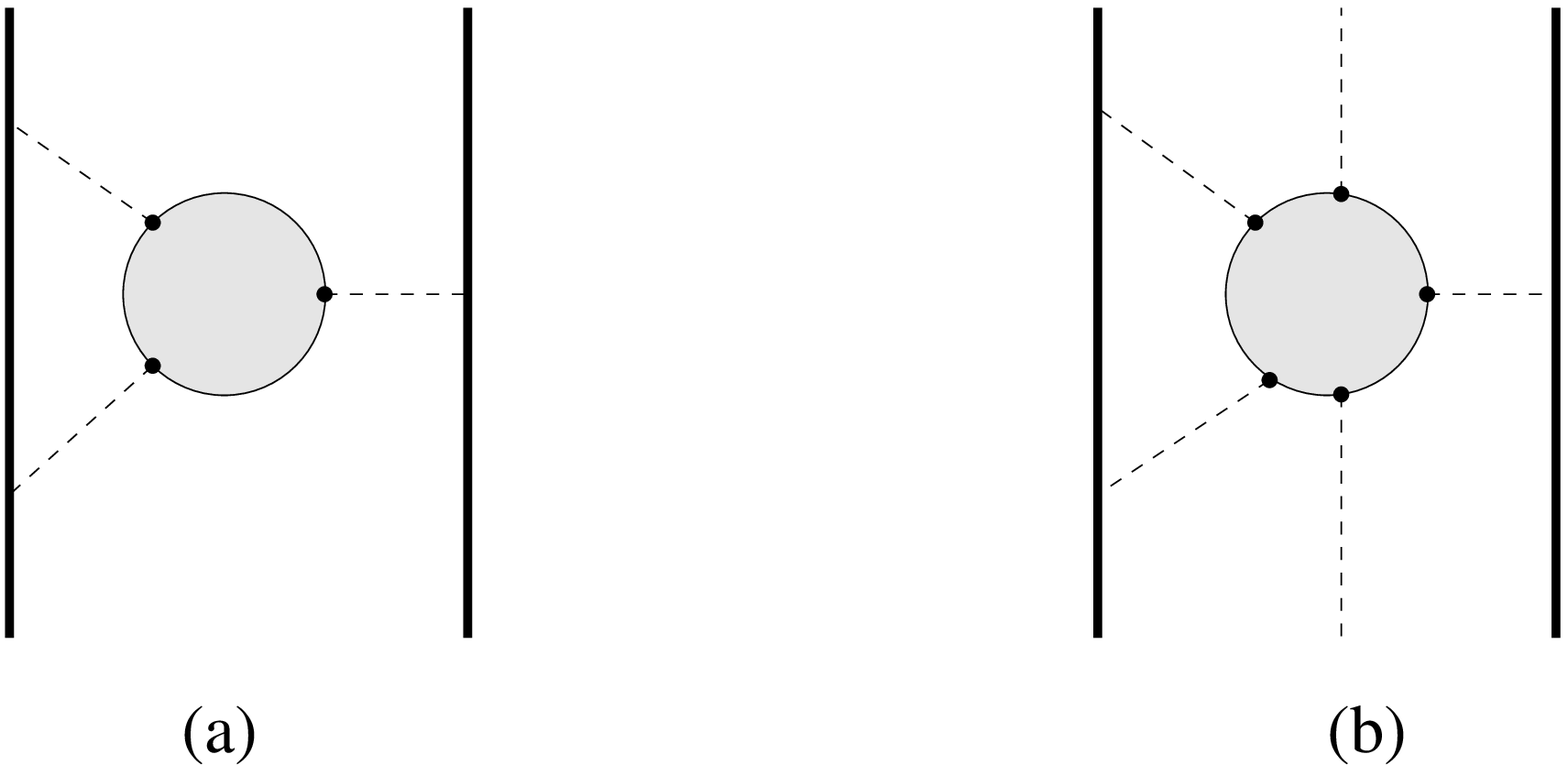}{1.5} 

\subsec{The bulk-boundary propagator}

We are interested in determining what data about the bulk spacetime
are contained in the CFT correlators in \gkpwd .  The first step is to
determine the bulk-boundary propagator in lorentzian spacetimes.

\paragraph{Euclidean propagator: }  The euclidean bulk-boundary
propagator was defined in \Wittads\ to be a solution of the bulk wave
equation that approaches a delta function at the boundary.\foot{In
practice, calculations are carried out by cutting off EAdS at
$z=\epsilon$ and removing this regulator at the end.   In order to
obey the CFT Ward identities we should actually require $\gbdel$ to
approach a delta function at $z = \epsilon$ \FMMR .}   It is more
convenient for us to Fourier transform with respect to the boundary
coordinates $\b$.   Then, $\gbdel (\k,\x)$ must be a solution to the
bulk wave equation that is non-singular in the bulk and is asymptotic
to a plane wave at the  
$z=0$ boundary.  This gives \refs{\FMMR, \BKLT}: 
\eqn\eucbb{
\gbdel (\k,\x) = A \, z^{d/2} \, K_\nu (qz) \, e^{-i \omega t + i
\vec{k} \cdot \vec{x}}\ .
}
Here $A$ is a normalization factor, $\k \equiv \{\omega, \vec{k} \}$
and $q^2 = \omega^2 + \vec{k}^2$.    The Bessel function $K_\nu$
vanishes exponentially as $z \rightarrow \infty$ and scales as
$z^{-\nu}$ at the boundary as $z \rightarrow 0$.

\paragraph{Poincar\'e propagator: }  We arrive at the Poincar\'e patch (PAdS) of CAdS by performing the Wick rotation $t \rightarrow it$ in \eads :
\eqn\pads{
ds^2 = {R^2 \over z^2} (-dt^2 + d\vec{x}^2 + dz^2 )
}
Fig.~3a displays a diametric slice of CAdS as an infinite tower of PAdS
patches.\foot{See the appendix of \BKL\ for more
details.}  Poincar\'e observers in each patch see past and 
future horizons ($H^-, \, H^+$) at $z=\infty$ where the patches meet.  The
boundary $B$ at $z=0$ of a given patch is conformal to the Minkowski
plane.  The PAdS/CFT 
duality relates physics within the Poincar\'e patch to a CFT defined
on the planar boundary.\foot{A CFT strictly speaking cannot be defined
on the Minkowski plane.  An infinite stack of planes is needed to
realize global conformal transformations \luschermack .  This is dual
in spacetime to the realization of the global isometries of CAdS on an
infinite tower of Poincar\'e patches.}  The PAdS propagator appearing
in \gkpwd\ is then obtained by continuation from \eucbb .  Setting
$q^2 = \omega^2 - \vec{k}^2$ we find:
\eqn\gbbtrans{\gbdel(\k,\x) = \tilde{A} \, z^{d/2} \, 
H_\nu^{(1)}(qz) \,
e^{-i\omega t + i\vec{k}\cdot\vec{x}}\ .
}
Here $H_\nu$ is a Hankel function, $q = \sqrt{q^2}$ and for spacelike
momenta ($q^2 < 0$) we choose the branch $q = i \sqrt{|q^2|}$.  Then,
for timelike momenta and positive $\omega$, $\gbdel$ is purely ingoing
at the horizon while for $q^2 < 0$, $\gbdel$ vanishes exponentially at
the horizon.  For timelike momenta there is also a spectrum of
normalizable mode solutions proportional to $J_\nu(qz)$ \BKL .  Such
modes may be added to the bulk-boundary propagator, inducing an
outgoing component at the past horizon.  The propagator \gbbtrans\
corresponds to a choice of Hartle-Hawking vacuum as usual (see
e.g. \cilar) 
upon taking
a euclidean continuation.

\ifig{Fig.~3}{Constructing CAdS from PAds}{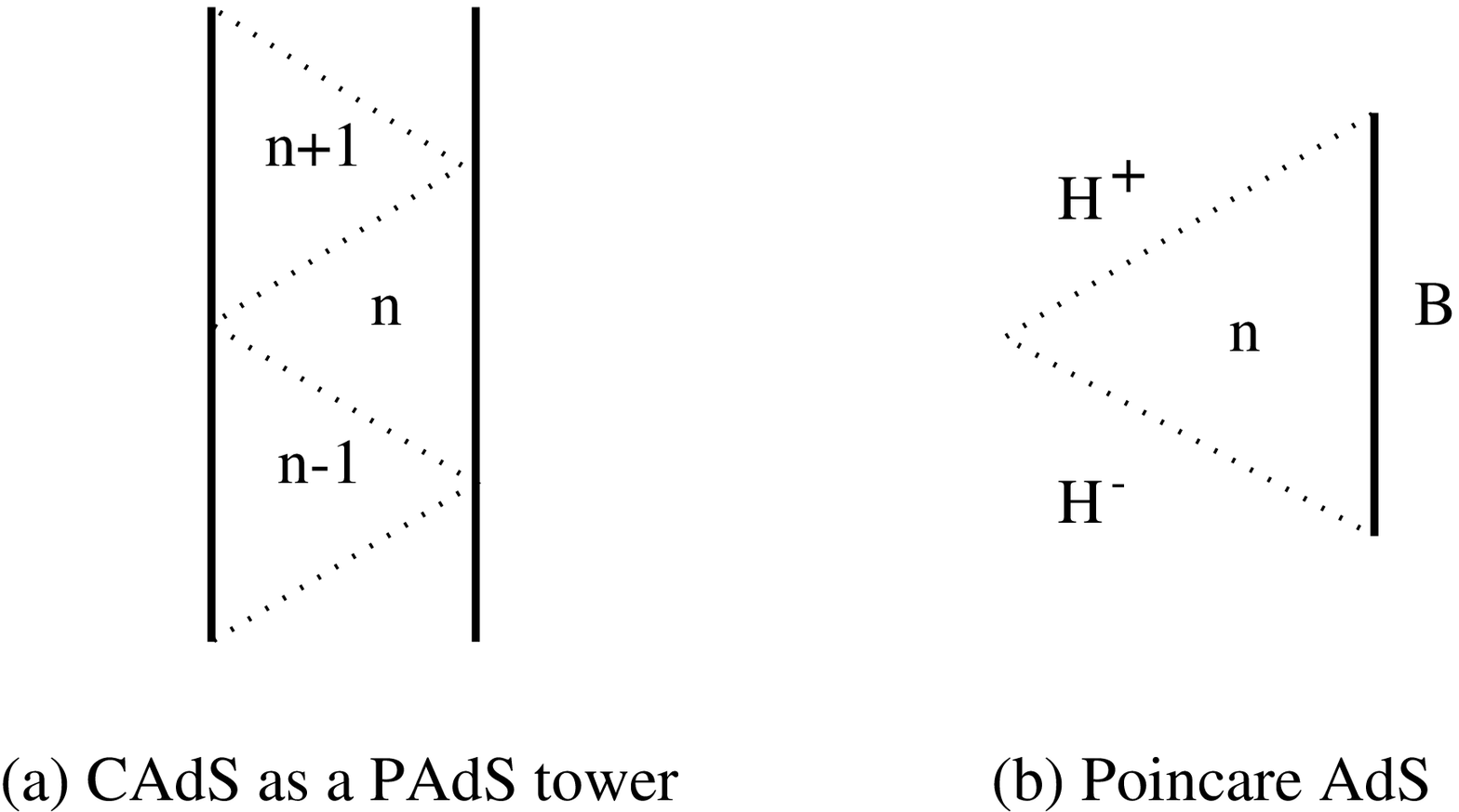}{2.0}

The PAdS bulk-boundary propagator may also be derived
by taking a scaling limit of the bulk Feynman propagator \BDHM : 
\eqn\bulktobb{
	G_{B\p}(\b;\b',z') = \lim_{z\to 0}
		z^{-2h_+} G_F(\b,z;\b',z')\ .
}
To show this we explicitly construct the bulk Feynman propagator from
the normalizable modes given in \BKL\ and find:\foot{Here it is
important to correctly normalize the bulk modes to $1$ in the
Klein-Gordon norm.} 
\eqn\bulkfeyn{
\tilde{G}_{B\partial}(\b_1;\b_2,z_2) =
\int_{q \geq 0} dq \, \left(\frac{q}{2}\right)^{1+\nu}
		\, \frac{1}{\Gamma(1+\nu)} \, 
	z_2^{d/2} \, J_\nu(qz_2) 
	\int \frac{d\vec{k}\, d\omega}{(2\pi)^d} \,
 		\frac{e^{-i\omega(t_1-t_2) + i\vec{k}\cdot
			(\vec{x}_1 - \vec{x}_2)}}
		{-\omega^2 + \vec{k}^2 + q^2 - i\epsilon}\ .
}
Fourier transforming and using Bessel function identities to convert
the integral over $q \geq 0$ into an integral over the real line
gives: 
\eqn\ftbbtwo{
	\tilde{G}_{B\partial}(\k;z,\b) =  
	{q^\nu \, z^{d/2}  \over 
	 2^{2+\nu} \, \Gamma(1+\nu) }
\, H_\nu^{(1)}(qz) \, e^{i\k\cdot \b} 
}
in agreement with \gbbtrans .  This relation between the bulk Feynman
propagator and the bulk-boundary propagator will be useful to us in
developing an S-matrix interpretation of CFT correlators.

\paragraph{CAdS propagator: }  Once again, in global AdS we seek a
solution to the wave equation approaching the delta function at the
boundary \refs{\Wittads,\FMMR}.   After expanding in spherical
harmonics on the boundary sphere we require a smooth solution with the
boundary behaviour 
\eqn\globallimit{
\gbdel \rightarrow
	e^{-i\omega \tau} Y_{l\vecm}(\Omega)(\cos\rho)^{2h_-}
}
as $\rho \rightarrow \pi/2$, where $Y_{l\vecm}$ are
the spherical harmonics (\cf\ \refs{\batemantwo,\absteg}). 
The unique solution for fixed
$\{\omega,l,\vec{m}\}$ is (see \eg\ \BKL): 
\eqn\globalprop{
\eqalign{
	\gbdel(\omega,l,\vecm;& \tau,\Omega,\rho) 
	= B \, e^{-i\omega \tau} \, Y_{l,\vecm}(\Omega) \cr
	&\times{(\cos\rho)^{2h_+} \, \, (\sin\rho)^l \, \,
		{}_2F_1\left(\half(2h_+ + l + \omega),
		\half(2h_+ + l - \omega), l + 
			\frac{d}{2},
		\sin^2 \rho \right)}\ ,
}}
where $B$ is a normalization constant.  For any given $l$, at the frequencies 
$\omega = 2h_+ + l + 2n$ with $n=\{0,1,\cdots\}$, 
there is a spectrum of normalizable modes that vanish at the boundary.  
These may be added to $\gbdel$ without changing the boundary behaviour. 
To eliminate the ambiguity, consider the euclidean continuation $\tau
\rightarrow i \tau$ of the CAdS metric \cads .   The bulk-boundary
propagator in that metric is a smooth solution to an elliptic
differential equation and is unique given a fixed boundary behaviour
\globallimit.   Continuing back to lorentzian signature does not
mix spherical harmonics or frequencies, giving \globalprop\ as the
propagator.

The CAdS propagator \globalprop\ has
singularities at the frequencies $\omega = 2h_+ + l + 2n$
corresponding to normalizable spacetime states.  This is a result
of the normalization condition \globallimit.
To see this, rewrite the
hypergeometric function ${}_2F_1$ as a function of $\cos^2\rho$
following \BKL:
\eqn\transfform{
	(\cos\rho)^{2h_+} \, (\sin\rho)^l \,F_1(\sin^2\rho) = 
	C^{+} \, \Phi^{(+)} + C^{-} \, \Phi^{(-)}\ ,
}
where $\Phi^{(+)}$ vanishes at the boundary and $\Phi^{(-)}$ achieves
the asymptotics \globallimit.   The coefficients $C^{\pm}$ are given
by:\foot{For $\nu\in \bbb Z$ the divergent Gamma function 
in the numerator of $C^{+}$ will be cancelled by a pole in the
denominator of $\Phi^{(+)}$.} 
\eqn\gammacoeff{
	C^{\pm} = \frac{\Gamma(l+\frac{d}{2})\Gamma(\mp\nu)}
		{\Gamma\left(\half(2h_\mp + l + \omega)\right)
		\Gamma\left(\half(2h_\mp + l - \omega)
			\right)}\ .
}
We maintain the asymptotics \globallimit\ by picking $B = 1/C_-$.
But $C_-$ vanishes when 
\eqn\normres{
	\omega  = \omega_{nl} = 
		2h_+ + l + 2n;\ \ \ \ n = 0,1,2,\cdots 
}
giving a divergence in $\gbdel$.  The divergence is proportional to
$\Phi^{(+)}$ which, at the magic frequencies \normres, are precisely
the normalizable states of the bulk theory.  In later sections, we
will interpret these singularities in terms of resonant scattering
behaviour. 

As in Poincar\'e coordinates, $\gbdel$ can be derived as a
limit of the bulk Feynman propagator in terms of the complete set of
modes derived in \eg\ \refs{\BKL}: 
\eqn\glmodes{\Phi^+_{nl\vecm}(\rho,\Omega) = N_{nl} \, (\cos
\rho)^{2 h_+} \, \, 
(\sin \rho)^l \, \, P_n^{(l-1+d/2,2 h_+-d/2)}(\cos 2\rho) \, \,
Y_{l\vecm}(\Omega)\ ,} 
where $N_{nl}$ is a normalization factor and 
$P_n^{k,m}$ are Jacobi polynomials.      
These modes have quantized frequencies 
$\omega_{nl}$ given in \normres.   The Feynman propagator is then:
\eqn\gfprop{G_B(\x_1, \x_2) = \langle T \, \Phi(\x_1)
\Phi(\x_2)
\rangle = \int_{-\infty}^{\infty} d\omega \sum_{n,l,\vecm}
{\Phi^{+*}_{nl\vecm}(\rho_1,\Omega_1)  \Phi^+_{nl\vecm}(\rho_2,\Omega_2)
e^{i\omega(\tau_1-\tau_2)}
\over \omega_{n,l}^2 - \omega^2 - i\epsilon}\ .}
We find the bulk-boundary propagator by taking one argument to the
boundary and rescaling:
\eqn\gbulkbdyd{\gbdel(t_1,\Omega_1; \x_2) = \lim_{\rho_1\rightarrow\pi/2}
(\cos\rho_1)^{-2h_+} G_B(\x_1,\x_2)\ .}
Projecting the resulting propagator onto definite frequencies and angular momenta gives:
\eqn\gproj{\gbdel(\omega,l,\vecm; \x) = \sum_n N'_{n,l} e^{-i\omega \tau} 
{\Phi^+_{nl\vecm}(\rho,\Omega)\over \omega_{nl}^2 - \omega^2 - i\epsilon}}
where $N'$ is a normalization factor.
Note that when $\omega\rightarrow \omega_{nl}$, the $n^{th}$ term in the sum 
dominates with an infinite coefficient.  We then get a pole factor times
$\Phi^+_{nl\vecm}$, replicating the resonant behaviour described above.

\newsec{Scattering from an AdS bubble}

We are now prepared to interpret the CFT correlators 
from the spacetime  perspective.   Begin 
by Fourier transforming the vacuum correlators 
of the CFT in \gkpwd.  Let $\psi_\alpha(\x)$ 
be the transformed bulk-boundary propagator with 
$\alpha \equiv \k = \{\omega, \vec{k} \}$ in 
Poincar\'e coordinates, and 
$\alpha \equiv \{\omega,l,\vec{m} \}$ in CAdS.   
Then \gkpwd\ becomes:
\eqn\boundcorr{
\langle T \, \CO_1(\alpha_1)
		\cdots\CO_n(\alpha_n) \rangle
	= \int \prod_{k=1}^{n} d\x_k \,  
\psi_{\alpha_1}(\x_1) \cdots 
		\psi_{\alpha_n}(\x_n)
\, 
\langle T \, \phi(\x_1)\cdots
		\phi(\x_n) \rangle_{{\rm T}}\ .
}
The left hand side of this expression is simply the Fourier transform
of a CFT correlator.  Here $\psi_\alpha$ is given by the mode solutions
\gbbtrans\ and \globalprop\ in PAdS and CAdS respectively.   So the
right side of \boundcorr\ looks just like an LSZ formula for an
S-matrix element: it is a truncated bulk Green function, with its legs
projected onto on-shell wavefunctions.  

Nevertheless, \boundcorr\ is not giving us a conventional LSZ
prescription for AdS states.   As observed in Sec.~2,
the notion of asymptotic states in CAdS in problematic because of the
reconvergence of geodesics.  We will make this observation precise in
Sec.~4 and show that conventional transition amplitudes may only be
defined for suitable trucations of CAdS.   Regardless, the modes
$\psi_\alpha$ appearing in \boundcorr\ are simply not the states of
the theory.  For general $\omega$ these modes have infinite action and
do not fluctuate, a property related to their divergence near the
infinite volume AdS boundary \BKL.\foot{Actually, in CAdS
$\psi_\alpha$ is normalizable at the magic frequencies \normres, but,
as discussed in Sec.2.1, it has a divergent coefficient and cannot be
thought of as an AdS state.} 

The expansion of the bulk-boundary propagator in terms of normalizable
modes  provides an important clue to the spacetime interpretation of
\boundcorr.    We will argue in Sec.~4 that 
a disturbance created at the AdS boundary at a given time can be
resolved into a sum of normalizable modes at a later time.  This is
dual to a statement that propagating states are produced in the CFT by
operators acting at early times.  In Sec.~4 we will use this fact to
interpret \boundcorr\ in terms of a transition amplitude 
for states in a temporal
AdS ``box" (a temporally truncated anti-de Sitter patch).
 
Here we pursue a complementary interpretation in terms of a spatial
AdS bubble.  The bulk-boundary propagator is explicitly constructed to
transport the influence of disturbances of the AdS boundary into the
interior of the bubble.  So \boundcorr\ summarizes the 
response of the AdS bubble to measurements by 
an observer external to the bubble.
More generally, imagine a large bubble
of AdS space inside an asymptotically flat spacetime.  (The bubble
might be metastable and could eventually dissipate.)  A strong form of
the holographic proposal says that quantum gravity inside any volume
should be described by a theory living on the boundary of the volume
\holog.  We then expect the large AdS bubble to be described by a CFT
living on its boundary.  An experimentalist probing the bubble can
make widely separated wavepackets and focus them to enter the bubble,
where they interact.  The S-matrix for this scattering process should
be described by correlation functions of the CFT.  Of course, the
fields would have to be suitably normalized to reflect the probability
that the wavepackets penetrate the AdS region.  But the non-trivial
part of the scattering process is precisely summarized by \boundcorr.
We will develop examples of this ``bubble" interpretation, displayed in
Fig.~4, in the remainder of this section.

\ifig{Fig.~4}{Scattering from an AdS bubble; here the dark lines denote a
finite distance boundary}{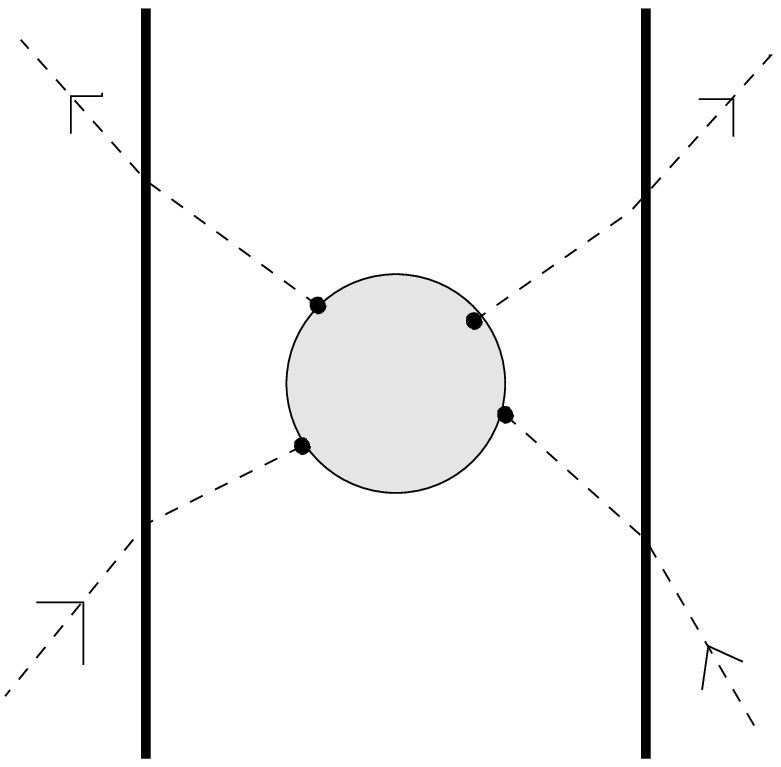}{1.5}

\subsec{S-Matrix: Poincar\'e coordinates}
In Poincar\'e coordinates for $\ads{5}$, a convenient example of a
bubble geometry is provided by the asymptotically flat 3-brane.  This
spacetime has an asymptotically flat region that patches onto an AdS
throat.  The duality of the bubble to a CFT accounts for the success
of CFT computations of supergravity scattering from 3-branes.  This
interpretation has been extensively explored in the work on brane and
black hole greybody factors (e.g., \greybody) and led directly to the
formulation of the AdS/CFT correspondence in \gkp.  So we will only
review the outlines of the argument here, and a pose a puzzle
regarding modes with spacelike momenta.

The asymptotically flat, extremal 3-brane has a metric
\eqn\threebrane{
ds^2 = H^{-1/2} \, (-dt^2 + d\vec{x}^2 ) + H^{1/2} \, (dr^2 + r^2 \,
d\Omega_5^2 ) 
}
with $H = 1 + R^4/r^4$,  $R^4 = 4\pi g \alpha^{\prime 2} N$ and a horizon
at $r=0$.  When $r \ll R$ we can set $H \approx R^4/r^4$.  Then taking
$z = R^2/r$ as the new radial coordinate yields Poincar\'e $\ads{5}$.
Alternatively, when $r \gg R$, $H \approx 1$ giving flat space. 

Consider a massless scalar $\phi$ satisfying the free wave equation in
this background.   Supergravity calculations of the scattering of this
scalar from the AdS region will be valid when the 3-brane scale is
large and incident energy small compared to the string scale.    This
was formalized in a ``double scaling limit" as
\refs{\klebnon,\klebrev}: 
\eqn\doubscal{
R^4/\alpha^{\prime2} \sim gN \rightarrow \infty ~~~~~~;~~~~~~
\omega^2\alpha^\prime \rightarrow 0\ .
}
Then setting $\phi = e^{i\vec{k} \cdot \vec{x} - i \omega t} \,
Y_{lm}(\Omega) \,\psi$ gives: 
\eqn\waveeq{
\left[ \left( 1 + {R^4 \over r^4} \right) (\omega^2 - \vec{k}^2) -
{l(l+4) \over r^2 } + {1 \over r^5} \partial_r r^5 \partial_r \right]
\psi = 0\ .
}
Euclidean AdS enjoys the same wave equation with $\omega^2$ replaced by 
$-\omega^2$.

\paragraph{Timelike momenta: }   When $q^2 = \omega^2 - \vec{k}^2 > 0$
solutions in the near and far region are Bessel functions.  We want a
scattering solution that is pure infalling at the brane horizon at
$r=0$ (or $z=\infty$ in the inverted AdS coordinate) and is smooth in
the far region.   So we choose: 
\eqn\timesol{
{\rm near: }~ \psi \propto z^2 \, H_{l+2}^{(1)}(qz)~~~~~~~;~~~~~~~
{\rm far:  }~ \psi \propto r^{-2} \, J_{l+2}(qr)
}
with $z = R^2/r$ and match these solutions at $r \approx R$.\foot{In
fact, the wave equation can be solved exactly in the 3-brane
background in terms of Mathieu functions \steveaki.}  The
bulk-boundary propagator in \gbbtrans\ now appears as the infalling
disturbance created by a probe from the asymptotic region.   The CFT
correlators in \boundcorr\ then describe the non-trivial interaction
of these disturbances in the AdS region.   Pure
ingoing boundary conditions are also required for a smooth euclidean
continuation --- $H_\nu^{(1)}$ continues to $K_\nu$ which vanishes
exponentially at the horizon while the other solution diverges.   This
is tantamount to choosing Hartle-Hawking boundary conditions for the
black 3-brane.  Different choices of bulk-boundary propagator for
AdS/CFT would correspond in the AdS bubble interpretation to different
boundary conditions for the 3-brane. 

\paragraph{Spacelike momenta: } The propagator \gbbtrans\ is defined
even for spacelike momenta with $q^2 = \omega^2 - \vec{k}^2 < 0$.  In
this case, setting $q = \sqrt{|q^2|}$, and demanding regularity at the
horizon and at infinity gives the near and far solutions: 
\eqn\spacesol{
{\rm near: }~ \psi \propto z^2 \, K_{l+2}(qz)~~~~~~~;~~~~~~~
{\rm far:  }~ \psi \propto r^{-2} \, K_{l+2}(qr)\ .
}
The solutions vanish exponentially both at the horizon ($z=\infty$)
and at infinity ($r=\infty$), and  so are not scattering states.   The
interpretation of \boundcorr\ for spacelike momenta therefore remains
puzzling.

\subsec{S-Matrix: global coordinates}

In order to interpret \boundcorr\ in global coordinates we want to
construct a CAdS bubble inside a spacetime that admits propagating
asymptotic states in a single asymptotic region.
A stable string solution of this kind is not
known.\foot{Of course the full analytic continuation
of the 3-brane metric has a CAdS interior and
an infinite number of asymptotic regions~\refs{\ght} whose interpretation can be
problematic.}
Nevertheless, for purposes of illustration,  consider a
spacetime with the metric:\foot{This simple metric is 
{\it not} asymptotically flat, although it is possible to write more
complicated ones that are.    Nevertheless, it permits the
definition of asymptotic particle states.} 
\eqn\adsbubmet{
\eqalign{
r < r_b: & ~~~~ 
ds^2= -(1+r^2)d\tau^2 + {dr^2 \over 1+r^2} + r^2 d\Omega^2  \cr 
r > r_b: & ~~~~
ds^2 = - (1+r_b^2) d\tau^2 + {dr^2 \over 1+r_b^2} +r^2 d\Omega^2 
\ .}
}
The interior metric is CAdS
in coordinates $\sec^2\rho = 1 + r^2$ and we have set the AdS scale
$R$ to $1$.  The induced metrics match on the surface $r = r_b$. 
This metric is not
expected to be a solution to string theory, and initial data
corresponding to such a geometry would certainly evolve with time.
Nevertheless, {\it any} string solution containing a metastable AdS
bubble will share certain features of scalar field propagation that we
now study. 

We want to compute the effective potential seen by a massless scalar
of angular momentum $l$ propagating in \adsbubmet.   It is useful to
work in terms of the tortoise coordinate $\rst$ satisfying $d\rst/dr =
\sqrt{-g_{rr}/g_{\tau\tau}}$.   Let us expand the scalar $\phi$ in
spherical harmonics as $\phi = Y_{l\vecm}\, u/R^{3/2}$ where $R^2$ is the
coefficient of $d\Omega^2$ in \adsbubmet.  Then radial motion is
governed by the action: 
\eqn\effact{S_{eff}\propto\int d\tau \, d\rst \left[ (\partial_\tau u)^2
- (\partial_\rst
u)^2 - V(\rst) u^2\right] }
with effective potential 
\eqn\effpot{V(\rst) =
-g_{\tau\tau}{l(l+1)\over R^2} + {\partial_\rst^2 R^{3/2}\over R^{3/2}}\ .  }
For $r<r_b$, the effective potential is that of Anti-de Sitter space, and
for $r\gg r_b$ 
\eqn\vdie{ V\sim {1\over \rst^2}\ ,}
allowing a definition of asymptotic states.  

Consider an incident wave moving towards $r=0$ from $r\gg r_b$.  Within
the AdS region we want a solution that is smooth at $r=0$.  This
selects the mode \globalprop.  Matching to the exterior incident wave
would give an absorption coefficient.  We are interested in a somewhat
different analysis where we construct a number of well separated
wavepackets in the asymptotic region and focus them to meet and
interact within the AdS bubble.  The actual interaction within AdS is
then completely described by the CFT correlators in \boundcorr.  To
compare to the full S-matrix computed in the asymptotic region we need
the probability of penetrating into AdS, which will provide the
correct normalization of CFT fields.  The latter data depend on the
precise patching of the AdS bubble into the asymptotic region.  The
CFT correlators in \boundcorr\ summarize the universal, and
nontrivial, part of the scattering amplitude which we can isolate by
taking $r_b \rightarrow \infty$ in \adsbubmet.  In this limit, we
recover CAdS.  So, although the modes \globalprop\ are not normalizable
and fluctuating within the full CAdS, the expression
\boundcorr\ is a universal limiting S-matrix for scattering from a
large CAdS bubble. 

This S-matrix interpretation also explains the origin of the
singularities of the bulk-boundary propagator at the magic frequencies
\normres.   As discussed in Sec.~2, there is a spectrum of normalizable
states in AdS at the frequencies $\omega_{nl} = 2h_+ + l + n$.   While
a wave incident on a CAdS bubble at generic frequencies will mostly
reflect, near resonance ($\omega \approx \omega_{nl}$) the
wavefunction for $r<r_b$ will scale as $1/(\omega - \omega_{nl} -
i\Gamma)$.  Here $\Gamma$ is a decay width for states inside the
bubble and $\Gamma \rightarrow 0$ as $r_b \rightarrow \infty$.    The
peak in the scattering amplitude at resonance becomes a pole as the
CAdS states become stable in the $r_b \rightarrow \infty$ limit.  This
explains the origin of the otherwise disturbing singularity in the
bulk-boundary propagator at the magic frequencies. 
 
\subsec{Interpretation}

We should stress that the bubble construction used to 
interpret \boundcorr\  is simply a useful artifice that can be
removed.  The non-trivial physics summarized in \boundcorr\ is
precisely that of scattering inside AdS.  An alternate way to see this
is to consider introducing sources for particles at finite but large
radius in AdS.  In the limit as the source goes to the boundary, and
at the same time has its amplitude scaled up to compensate for the
vanishing tunneling factors, we also recover \boundcorr. 

Furthermore, our discussion has used a diagrammatic expansion for the
nearly free limit which is suspect in the fully holographic theory.
Nonetheless, our interpretation of \boundcorr\ as a natural analogue
of an S-matrix can be expected to hold in the full theory,  providing
a path from boundary to bulk physics.

\newsec{States and transition amplitudes}

In the previous section, we have found an interpretation
for truncated time-ordered bulk Green functions convolved against
non-normalizable modes.   We might expect that Green
functions convolved against normalizable modes have a meaning as
well. Indeed,
the conventional LSZ
prescription in flat space 
relates S-matrix elements, defined as the overlap of
``in'' and ``out'' states with well-defined particle number, to
truncated Green functions convolved against one-particle
wavefunctions.  Therefore, in this
section we return to the issue
of fluctuating states in AdS and transition amplitudes between
them.  We will show in Sec.~4.1 that this prescription  fails
in CAdS space because ``in'' and ``out'' states, propagating  as a
collection of widely separated single non-interacting particles, do
not exist.  Roughly this is because lightlike geodesics hit the
boundary in finite time, while spacelike geodesics always reflect
back into the bulk after finite time.  So
particles (classical or quantum) do not become infinitely separated in
the far past.    To avoid this subtlety we will partition AdS into
patches within which geodesics do not reconverge, and show that the
appropriate CFT dual computes transition amplitudes between
fluctuating states defined on the patch. 

\subsec{The failure of LSZ in CAdS}

To define ``in" and ``out" states in the usual manner we follow
Sec.~3.1 of \weinbergone.   Assume we can split up the Hamiltonian as:

\eqn\fullhamilt{
	H= H_0 + H_I\ ,
}
where $H_0$ is a free Hamiltonian, but contains the renormalized
masses so that $H_0$ and $H$ have the same one-particle spectrum.   Then
in Heisenberg picture, both free and interacting one-particle states
satisfy: 
\eqn\states{
	H_0 \Psi_{0,\alpha} = E_\alpha \Psi_{0,\alpha} ~~~~~~;~~~~~~
	H \Psi_{\alpha}  = E_\alpha \Psi_{\alpha}\ .
}
Here $\alpha$ denotes the full set of quantum numbers -- momenta,
spin, and so on.  ``In'' states $\psi_\alpha^+$ are defined by the
requirement that wavepackets constructed from them approach free
wavepackets at early times:
\eqn\infielddef{
e^{-iHt} 	\int d\alpha \, g(\alpha) \, \Psi^+_\alpha \
{\buildrel t\to -\infty  \over \longrightarrow } \
e^{-iH_0 t} \int d\alpha \, g(\alpha) \, \Psi_{0,\alpha}
}
Here $g(\alpha)$ is a kernel used to define wavepackets. 
``Out'' fields $\Psi^-_\alpha$ are defined identically,
with the limit replaced by $t\to\infty$.

The Lippmann-Schwinger equation provides a recursive solution:
\eqn\lippschwing{
	\Psi^\pm_\alpha = \Psi_{0,\alpha}
		+ \int d\beta \, \, \frac{\bra{\Psi_{0,\beta}}
			H_I \ket{\Psi^\pm_\alpha}}
			{E_\alpha - E_\beta \pm i\epsilon} \, \, 
		\Psi_{0,\beta}\ 
}
Now we multiply both sides by $g(\alpha) e^{-iE_\alpha t}$ and sum
over $\alpha$.
Wave packets built from \lippschwing\ satisfy \infielddef\ if:
\eqn\interpiece{
	\lim_{t\to\mp\infty}
	\int\int d\alpha \, d\beta \, e^{-iE_\alpha t} \,
		g(\alpha) \, \, \frac{\bra{\Psi_{0,\beta}}
			H_I \ket{\Psi^\pm_\alpha}}
			{E_\alpha - E_\beta \pm i\epsilon}
		\, \, \Psi_{0,\beta} = 0\ .
}
We will check this condition in Minkowski and CAdS spaces.

\paragraph{Minkowski spacetime: } In Minkowski space, the energy
spectrum is continuous.  Extending the integral in \interpiece\ over
positive and negative energies and integrating over $\alpha$, the
contour is closed in the upper half-plane for ``in" states.  The pole
in the denominator is then avoided.   There may be additional poles in
$g(\alpha)$ controlling the energy width of the state, and poles in
the matrix element controlling the duration of collision.  For large,
negative $t$, the integral will be a sum over these poles and
exponentially damped.  ``Out" states may be treated similarly

\paragraph{CAdS spacetime:} In CAdS the energies of particles are
discrete and given by \normres.   This spectrum is essentially
dictated by the representation theory of the AdS isometry group; so we
do not expect the integral spacing of levels to be affected by quantum
corrections. Then $k$ particles with masses $m_i$ have a total
energy,\foot{Generally, the multiparticle energy may not be a sum
of single particle energies; this formula is for illustration's sake.}
\eqn\kparticleenergy{
	E = \sum_{i=1}^k
	\left( 2h_+^{(k)} + l_k + 2n_k \right)\ , 
}
and the left side of \interpiece\ is almost periodic in time with
period $2\pi$, up to an overall phase
$$	e^{-i \left(\sum_{i=1}^k 2h^{(k)}_+\right)t}\ .$$
Thus the last term in 
\lippschwing, summed against $g(\alpha)\,e^{-iE_\alpha t}$,
is finite and oscillatory in time, and the conditions in
\infielddef\ cannot be met.  This almost-periodic behavior
of wavefunctions is the quantum reflection of the
classical statement that geodesics are periodic in global time.

\subsec{Transitions between states}
\def\pads#1{{\rm PAdS}_{#1}}

Despite the subtlety described above we might expect to construct
physically interesting
transition amplitudes between states within a patch of AdS in which
geodesics are not periodic.  A particularly convenient patch is
provided by the Poincar\'e coordinates discussed in Sec.~2.  The CAdS
and PAdS metrics were presented in Sec.~2 and, as shown in Fig.~3,
the global spacetime can be constructed as a PAdS ladder.  We will
refer to the nth patch as $\pads{n}$.  Each patch has a timelike
boundary $B_n$ and a past ($H_n^-$)and future ($H_n^+$) horizon.

\paragraph{CAdS from PAdS: } To define the theory on $\pads{n}$ we
need boundary conditions at $B_n$, $H_n^+$ and $H_n^-$.  Boundary data
on $B_n$ are specified by turning on frozen, non-normalizable
solutions of the form \refs{\FMMR,\BKL,\BKLT}:
\eqn\nonnormp{
\psi_{\k}(\x) = A \, z^{d/2} \, H_\nu^{(1)}(qz) \, e^{-i\omega t +
i \vec{k}\cdot\vec{x}}\ .
}
For timelike frequenices($q^2 = \omega^2 - \vec{k}^2 > 0$), these
modes are pure ingoing (outgoing) at the horizon for $\omega > 0$
($\omega < 0$),
\eqn\nonnasymp{
\psi_{\k}(\x) \, {\buildrel z\to \infty \over \longrightarrow} \, 
B \, z^{(d-1)/2}\, e^{i (q z - \omega t) + i \vec{k} \cdot\vec{x} }\ .
}
Modes with spacelike frequencies ($q^2 = \omega^2 - \vec{k}^2 < 0$)
vanish exponentially at the horizon and are not of interest to us
here.  The normalizable mode solutions of the wave equation are
\refs{\breitfreed,\avisetal,\BKL}:
\eqn\normp{
\phi_{\k}(\x) = A \, z^{d/2} \, J_{\nu}(qz) \, e^{-i\omega t + i \k \cdot \x}
} 
and are a mixture of outgoing and ingoing modes at the horizon 
\eqn\normasymp{
\phi_{\k}^1(\x) \, {\buildrel z\to \infty \over \longrightarrow} \, 
B \, z^{(d-1)/2} \, \left[ e^{i (q z - \omega t- \frac{\pi\nu}{2} -
 \frac{\pi}{4} ) + 
 i \vec{k} \cdot\vec{x} }
+ e^{-i (q z + \omega t- \frac{\pi\nu}{2} - \frac{\pi}{4} ) + i
 \vec{k} \cdot\vec{x}} \right]\ . 
}
We can patch together these modes on a series of  patches to obtain a
solution to the CAdS wave equation.  For example, suppose that the
mode $\phi_{\k}$ is  present on $\pads{n}$.   To match the flux at the
horizon we could turn on $\psi_{\k}$ with positive frequency on
$\pads{n-1}$ and $\psi_{\k}$ with negative frequency on
$\pads{n+1}$. A general solution on CAdS which vanishes at very early
and very late times is constructed by turning on a collection of
$\psi$ with positive frequency on some early patch, matching onto a
sequence of normalizable and non-normalizable modes in later patches, and
then soaking up the flux by a collection of  $\psi$ with negative
frequency on a late patch.   This process of splicing solutions is
displayed schematically in Fig.~5.

We have just described how to constructed classical solutions to the
CAdS wave equations by sewing PAdS modes together.  In the classical
limit, given Dirichlet boundary conditions for $\pads{n}$, the
normalizable modes \normp\ propagate undisturbed from the past to the
future horizon.  In the interacting theory, these modes are the
candidate early and late time states between which we wish to compute
transition amplitudes.  In the following, we will describe how to do
this from the dual CFT perspective.

\ifig{Fig.~5}{CAdS solutions from PAdS}{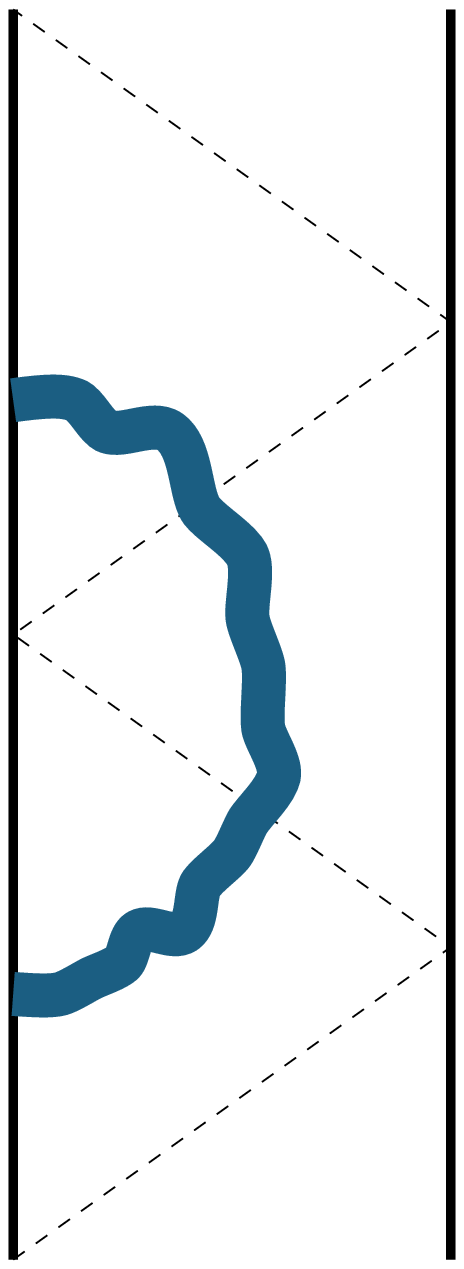}{1.7}

\paragraph{CFT dual: } The construction of CAdS as a tower of PAdS
patches is dual to the definition of lorentzian CFTs on a stack of
Minkowski diamonds \luschermack.   The classical limit in spacetime is
dual to the large $N$ limit of the CFT, and the splicing prescriptions
for classical solutions translate into relations between the CFTs
defined on the boundaries $B_n$ of the tower of patches.   Let ${\cal O}_n$
be the CFT operator on $B_n$ that is dual to a field $\phi$.
Deforming the large $N$ CFT on $B_{n-1}$ by adding the term
$\int_{\partial} \phi_0(\b) \,  {\cal O}_{n-1}(\b)$ to the lagrangian is dual
in spacetime to a classical non-normalizable mode that approaches
$\phi_0(\b)$ on the boundary \refs{\gkp,\Wittads,\BKL}.   Working in a
Fourier basis, a source for ${\cal O}_{n-1}(\k)$ in the CFT is therefore dual
to turning on the bulk-boundary propagator \gbbtrans\ (or $\psi_{\k}$)
as a classical mode in the bulk.  The patching conditions for CAdS
then induce a classical mode in $\pads{n}$ to soak up the incoming
flux at the past horizon.  Choosing the normalizable mode $\phi_{\k}$
is dual to placing the CFT on $B_{n}$ in the corresponding
``coherent'' state \BKLT.  Finally, we can match onto a negative
frequency $\psi_{\k}$ mode in $\pads{n+1}$.   This naturally provides
a source term for a negative frequency mode of  ${\cal O}_{n+1}$. 

From this we learn that states in $\pads{n}$ can be created
(annihilated) by positive (negative) frequency operators ${\cal O}$
acting on past (future) patches.  We are interested in transitions
between states of $\pads{n}$.  So it is sufficient to consider
operators acting on $B_{n-1}$ and $B_{n+1}$.   Writing the positive and
negative frequency parts of $\CO$ as $\CO^+$ and $\CO^-$,
incoming states $|s\rangle_n$ at $H_n^-$ are defined as: 
\eqn\instate{
|s\rangle_n \equiv
{\cal O}^+_{n-1}(\k_1) \cdots {\cal O}^+_{n-1}(\k_s) |0\rangle_{n-1}\ .
}
Here the incoming state at the past horizon of the nth patch has been
identified with the action of positive frequency operators on the Poincar\'e
vacuum state of $B_{n-1}$.   Likewise, the outgoing state at the
future horizon can be written as:  
\eqn\outstate{
{}_n\langle s^\prime | \equiv
{}_{n+1}\langle 0 | {\cal O}^-_{n+1}(\k_s) \cdots {\cal
O}^-_{n+1}(\k_1)\ .
}
Here the action of negative frequency operators on the vacuum of
$B_{n+1}$ has been identified with the outgoing state on $\pads{n}$.
In the nearly free limit, this associates the normalizable modes
\normp\ with the in and out states of PAdS.  But \instate\ and
\outstate\ have the added virtue of being well-defined even when
interactions are turned on at finite $N$.  What is more, these
definitions accord well with our intuition that states in a CFT are
created and annihilated by operators acting at early and late times.
Here, operators on $B_{n-1}$ and $B_{n+1}$ are acting before and after
the beginning and end of time from the $\pads{n}$ perspective.

\def\y{{\bf y}}

\paragraph{PAdS transition amplitudes: } We finally have all the
ingredients to assemble transition amplitudes in $\pads{n}$ from the
CFT perspective.   Transition amplitudes in $\pads{n}$ are defined as the
overlaps $\langle s^\prime| s \rangle$.   From the defintion of the in
and out states, the calculation we must perform is: 
\eqn\scalc{
\langle s^\prime| s \rangle = \langle 0 | \prod_i {\cal O}_{n+1}^{-}(\k_i)
\, \prod_j {\cal O}_{n-1}^{+}(\k_j) | 0 \rangle\ .
}
The correlation functions are computed on the  CAdS cylinder where the
CFT  is  actually defined using the master formula \gkpwd.  The
notation ${\cal O}_{m}(\k)$ indicates the Fourier mode of an operator which
has support only on the patch $B_{m}$ of the CAdS boundary cylinder.
So, following \boundcorr, after Fourier transforming we compute: 
\eqn\smear{
\langle s^\prime | s \rangle
	= \int \prod_{i=1}^k \prod_{j=1}^l \left(d\x_i \, d\y_j \,
             \psi^-_{\k_i}(\x_i)  \,
             \psi^+_{\k_j}(\y_j) \right) \,           
\langle T \phi(\x_1)\cdots \phi(\x_k) 
        \phi(\y_1) \cdots \psi(\y_l)
\rangle_{{\rm T}}\ .
}
Here $\x$ and $\y$ are CAdS coordinates while $\psi^{\pm}$ are
positive and negative frequency, non-normalizable Poincar\'e modes of
the form \nonnormp\ written in global coordinates.  We then interpret
\scalc\ as a transition amplitude for states on $\pads{n}$.
Diagramatically we obtain Figs.~6a and 6b where the truncation of CAdS
diagrams by the PAdS patch leaves legs intersecting the past and
future horizons.  These legs are wavefunctions representing the in and
out states and, in the nearly free limit, will be given in $\pads{n}$
by the normalizable AdS modes.  The shaded circles on the propagators from the
CAdS boundary to the PAdS horizons indicate that the full interacting
propagator should be used to define the meaning of in and out states
of $\pads{n}$.  In general this means that these states do
not have a clear interpretation as single particle wavefunctions, but
they nevertheless give a basis for the initial and final time boundary
conditions for PAdS.

\ifig{Fig.~6}{PAdS diagrams computing transition
amplitudes}{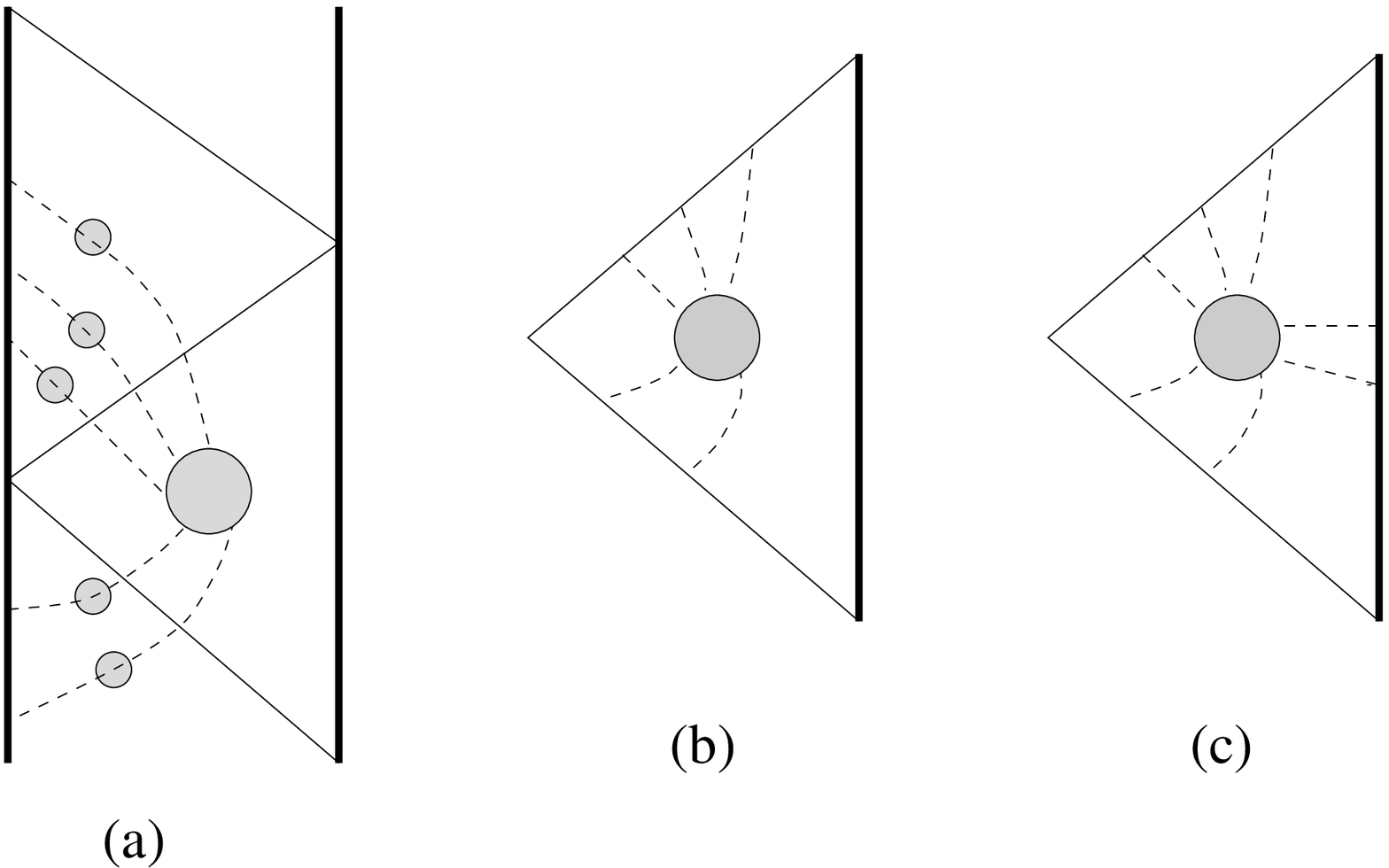}{2.5}  

In addition to the transition amplitudes \scalc\ we may compute
amplitudes like 
\eqn\smatscat{
\langle s^\prime | {\cal O}_1(\k_1) \cdots {\cal O}_n(\k_m) | s \rangle \ .
}
Diagramatically these are represented in Fig.~6c.  Following
previous sections these amplitudes may be interpreted as amplitudes
for scattering from an excited AdS bubble.   The local operator insertions on the $\pads{n}$ boundary appear in the bulk as non-normalizable modes while the states are represented by normalizable modes in the nearly-free limit.

\subsec{Discussion}

\def\tads#1{{\rm TAdS}_{#1}}

\paragraph{Conclusion and subtleties: } 
At first glance \smear\ appears to describe PAdS transition amplitudes as truncated bulk Green functions convolved again normalizable wavefunctions.  
This interpretation basically applies in the nearly free limit,
leading to Fig.~6a.   In general, however, the interactions in \smear\
may occur anywhere within CAdS, leading to Fig.~7.   One way of
dealing with this subtlety is to create states via well separated
wavepackets in $\pads{n-1}$  that do not interact until they have
entered $\pads{n}$.  This is certainly possible in the semiclassical
limit, and in this way the in-states on the past horizon remain under
reasonable control.   Another approach is to simply define the states
on the past (future) horizon to be the objects on those surfaces that
result via propagation through the previous (later) patches.  In the
latter approach, as illustrated in Fig.~7, the particle number of the
in and out states may be indefinite;  but these states are nonetheless well
defined via the operator constructions \instate\ and \outstate. 

\ifig{Fig.~7}{Contributions from the truncated Green function
outside $\pads{n}$.}{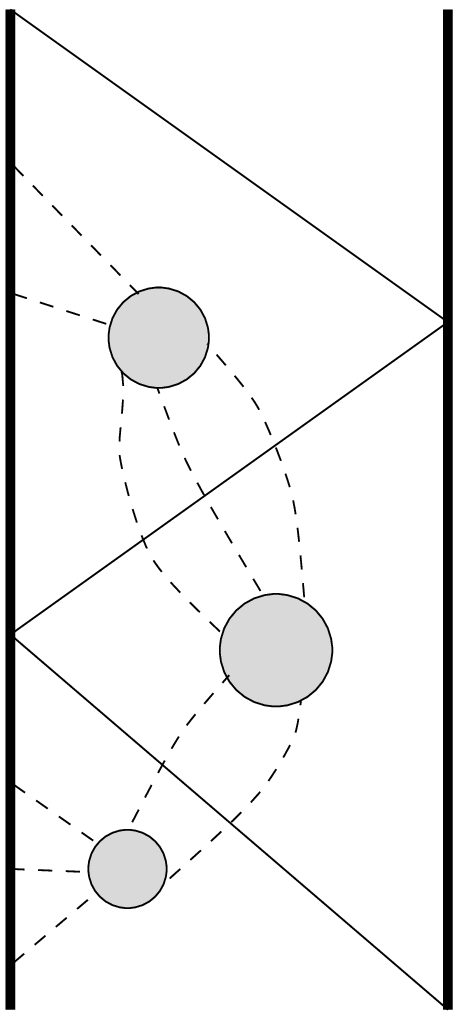}{2.5}

\paragraph{Truncated AdS: } Instead of building up CAdS as a sequence
of PAdS patches we could simply construct a sequence of cylinders of
length $T$ in global time.  We will call the nth cylinder $\tads{n}$.
As we have discussed, the effect of acting on the boundary with a CFT
operator is transported into spacetime by the bulk-boundary
propagator.  By definition, this propagator in position space is a
delta function on the boundary.  It follows from this that a
bulk-boundary propagator with support only on the $\tads{n-1}$
boundary must create a solution that is expandable in the normalizable
mode solutions of the next patch $\tads{n}$.  We can therefore
precisely mimic the PAdS construction above to construct transition
amplitudes for TAdS states.  Again, in the nearly free limit,
these are normalizable modes convolved
against truncated Green functions, all in patch $n$, but a more complicated interpretation applies in the fully interacting theory.

\paragraph{Flat space S-matrix from AdS: } The methods of this section
were designed to describe the scattering of wavepackets created on the
early and late boundaries of a patch.   In the very large N limit, the
spacetime is nearly flat in the interior.   Our techniques can readily
be used to to construct wavepackets within the flat region of the bulk
spacetime.  Then \scalc\ gives a holographic computation of transition
amplitudes between states in flat space.  A very similar logic has
been pursued recently in \refs{\joesmatrix,\lennysmatrix} where
equations like \boundcorr,\scalc\ were also interpreted as transition amplitudes.

\newsec{Towards bulk correlators}

It has been suggested that the on-shell
transition amplitudes recovered in the preceding
sections contain all the information we can 
extract from a holographic
theory \refs{\banksgreen}.  Nonetheless, we know
that approximate spacetime locality holds in nature and 
we would like to see how it is encoded in the CFT.

Eqs. \bulktobb,\euccorr\ imply a 
relation between bulk and boundary
correlators \BDHM:
\eqn\boundrestr{
	\langle \CO_1({\bf b}_1)\cdots\CO_n({\bf b}_n) \rangle
	= \prod_{k=1}^n \lim_{z_k\to 0} z_k^{-\Delta_{(k)}}
	\langle\phi(z_1,{\bf x}_1)\cdots\phi(z_n,{\bf x}_n)\rangle\ .
}
where $\Delta_{(k)} \equiv 2h_{+(k)}$ is the conformal dimension of $\CO_k$.
This is true for general correlators, suggesting the relation
\eqn\oprest{
	\CO({\bf b}) = \lim_{z\to 0} z^{-\Delta}\phi(z,{\bf x})\ .
}
This is also implied in the $N\to\infty$ limit by Eq.~(15) of \BKLT.
If an inverse of the map \oprest\ exists, we could reconstruct
off-shell bulk correlators from boundary data, and directly
investigate spacetime locality.  
This section will explore attempts to 
construct such an inverse.

\subsec{The free field map}

In the free limit we can invert \oprest\ by 
comparing the bulk and boundary mode 
expansions \refs{\BDHM,\BKLT}.    
The bulk expansions for a scalar 
field of mass $m$ are fixed 
by the equations of motion and canonical commutation 
relations in spacetime.\foot{The mass is assumed 
to include any curvature couplings.} 
On the boundary we simply expand in Fourier 
modes and fix the normalization via \oprest.   

\subsubsec{Mode expansions}

\paragraph{Global coordinates:}
The bulk mode expansion is: 
\eqn\gfreeop{
\eqalign{
	\Phi(t,\rho,\Omega) & = 
	\sum_{n,\ell,\vec{m}} \left(
		\frac{\Gamma(1+n)\Gamma(\Delta + \ell + n)}
		{\Gamma(\ell+\frac{d}{2} + n) 
			\Gamma(1+\nu+n)N_\ell} \right)^{\half}\times\cr
	&\ \ \ \ 
(\sin\rho)^\ell \, (\cos\rho)^{\Delta} \, 
	P_n^{(\ell + \frac{d}{2}-1,\nu)} (\cos2\rho) 
\times \left\{
		e^{-i\omega_{n\ell} t} \, Y_{\ell \vec{m}}(\Omega)
		\, \hat{a}_{n\ell \vec{m}} + 
 {\rm h.c.}
\right\} \ .
}}
(See \refs{\avisetal,\breitfreed,\meztown,\BKL} for a discussion of mode solutions.) 
The frequencies $\omega_{n\ell}$
are defined in Eq.~\normres\ and the spherical harmonics
$Y_{\ell\vecm}$ satisfy the standard orthonormality conditions
\eqn\ylmortho{
	\int d\Omega \, Y^\ast_{\ell\vecm}(\Omega) \, Y_{\ell' \vecm'}(\Omega)
		= \delta_{\ell\ell'}\, \delta_{\vecm\vecm'} \, N_{\ell}\ . 
}
Then the creation and annihilation operators obey commutation
relations
\eqn\freecr{	
	\left[\hat{a}_{n\ell \vec{m}}, \, 
	\hat{a}^\dagger_{n'\ell' \vec{m}'}\right]
		= 
	\delta_{nn'} \, \delta_{\ell\ell'} \, \delta_{\vec{m}\vec{m}'}\ .
}
The boundary expansion follows from 
Fourier expansion of the CFT operators and \oprest: 
\eqn\freeprim{
	\CO(t,\Omega) 
= \frac{1}{\Gamma(1+\nu)}
		\sum_{n,\ell,\vec{m}} \left(
	\frac{\Gamma(1+n+\nu)\Gamma(\Delta + \ell + n)}
		{N_\ell \Gamma(1+n)\Gamma(\ell+\frac{d}{2}+n)}
	\right)^\half \times \,
\left\{e^{-i\omega_{n\ell} t} Y_{\ell \vec{m}}(\Omega)
		\hat{a}_{n\ell \vec{m}}
      + {\rm h.c.} \right\}
}

\paragraph{Poincar\'e coordinates: } 
In PAdS $\Phi(z,b)$ can be expanded as
\eqn\pfreeop{
	\Phi(z,t,\vec{x}) 
= \int d^{d-1}\vec{k} \, dq \,
	z^{\frac{d}{2}} \, J_\nu (q z) \, \left[
	\frac{q}{2(2\pi)^{d-1}\omega(q,k)}\right]^{\half}\times
\left\{
		e^{-i\omega(q,k)t + i\vec{k}\cdot\vec{x}}
			\hat{b}_{q\vec{k}} +
     {\rm h.c.}
		\right\}\ .
}
Here $\omega(q,k) = q^2 - k^2$.  $J_\nu$
are Bessel functions, and we follow the conventions 
of \refs{\absteg,\batemantwo}.
The creation and annihilation operators $(\hat{b},\hat{b}^\dagger)$
satisfy the commutation relations
\eqn\pcr{
	[\hat{b}_{q\vec{k}},\hat{b}_{q' \vec{k}'}]
		= \delta(q-q') 
		\delta^{(d-1)}(\vec{k}-\vec{k}')\ .
}
Fourier expanding the dual operator and using \oprest\ 
gives:\foot{The $z\to 0$ limit is formal and   is taken
term by term in the expansion in $q$; otherwise
there is an order-of-limits issue in
using the small-argument asymptotics for $J_\nu$
at the upper end of the integral.}
\eqn\pfreeprim{
	\CO(t,\vec{x}) 
= \int dq \, d^{(d-1)}\vec{k} \, \left( \frac{q}{2}
		\right)^{\nu+\half} \frac{1}{\Gamma(1+\nu)
			\left[(2\pi)^{d-1}\omega(q,k)\right]^\half} \,
\times \, 
\left\{e^{-i\omega(q,k)t + i\vec{k}\cdot\vec{x}}
			\hat{b}_{q,\vec{k}} +
          {\rm h.c.}
		\right\}\ .
}

\subsubsec{Bulk operators from boundary operators?}

The bulk and boundary mode expansions can be conveniently related via
a ``transfer matrix'' \BDHM:
\eqn\transf{
	\phi(z,\b) = \int d\b' \, M(z,\b;\b') \, \CO(\b')\ .
}
The explicit transfer matrices are readily found
by Fourier transforming \freeprim\ and \pfreeprim.   In CAdS this
gives 
\eqn\globaltrans{
\eqalign{
	M(\rho_1,t_1,\Omega_1;t_2,\Omega_2) &=
	\sum_{n,\ell,\vec{m}} \frac{\Gamma(1+\nu)\Gamma(1+n)}
	{N_\ell \Gamma(1+n+\nu)}(\sin\rho_1)^\ell
	(\cos\rho_1)^{\Delta} 
	P_n^{(\ell+\frac{d}{2}-1,\nu)}(\cos 2\rho_1)\times\cr
	&\ \ \ \ \ \times\left\{
	e^{-i\omega(t_2-t_1)}Y^\ast_{\ell \vec{m}}(\Omega_1)
		Y_{\ell \vec{m}}(\Omega_2) +
	e^{i\omega(t_1-t_2)} Y_{\ell \vec{m}}(\Omega_1)
		Y^\ast_{\ell \vec{m}}(\Omega_2)\right\}\ ,
}}
and in Poincar\'e coordinates,
\eqn\pointrans{
\eqalign{
	M(z_1,t_1,\vec{x}_1;t_2,\vec{x}_2)
	&  = \Gamma(1+\nu) \, \int d^{d-1}\vec{k} \, dq \, 
  \left( {2 \over q} \right)^\nu
\, 
	z_1^{\frac{d}{2}}  \, J_\nu(qz_1) \, \times 
\cr
	&\ \ \ \ \ \times 
\left\{
	e^{-i\omega(q)(t_2 - t_1) + i\vec{k}\cdot(\vec{x}_2 - \vec{x}_1)}
      + e^{i\omega(q)(t_2 - t_1)  - i\vec{k}\cdot(\vec{x}_2 - \vec{x}_1)}
	\right\}\ .
}
}

This gives a concrete proposal for the map from boundary to bulk fields,
and, as in \BDHM, we might now attempt to promote it to the interacting
theory and infer local bulk correlators via
\eqn\bdhmeq{
	\langle \phi(z_1,\b_1) \phi(z_2,\b_2)\cdots
	\phi(z_n,\b_n)\rangle {\buildrel ? \over =} \int 
	\prod_{i=1}^n
	\left[d\b_i' M(z_i,\b_i;\b_i')\right]
	\langle \CO(\b_1')\CO(\b_2')\cdots\CO(\b_n')\rangle\ .}
However, as the authors of \BDHM\ noted, this cannot be correct
because the result satisfies the free wave equation rather than the
interacting Dyson-Schwinger equations.  The next subsection explains
this using the Lehmann representation for bulk fields.

In fact, from \boundcorr\ we know that the right hand side of \bdhmeq\
can be expressed as:
\eqn\cbdhm{\eqalign{
\int\prod_{i=1}^n&
	\left[d\b_i' M(z_i,\b_i;\b_i')\right]
	\langle \calo(\b_1')\calo(\b_2')\cdots\calo(\b_n')\rangle\cr
	&= \int \left[\prod_{k=1}^n d\x_k' \right]
	\prod_{k=1}^n\left[ d\alpha_k \, \phi_{\alpha_k}^*(\x_k) \,
	\psi_{\alpha_k}(\x_k') 
\right]\langle
	\phi_1(\x_1') \cdots\phi_n(\x_n') \rangle_T \ ,\cr}}
where $\phi_\alpha^*$ denotes a normalizable wavefunction and 
the integrals over $\alpha_k$ are shorthand for the full
sum/integral over Fourier conjugate variables appropriate to either the
global or Poincar\'e case.   Prior to integrating over $\alpha_k$, the
right side of \cbdhm\ is a truncated bulk Green function projected
onto non-normalizable, on-shell wavefunctions.   This is further
convolved against normalizable modes in \cbdhm.  So it is manifest
that the right side of \bdhmeq\ satisfies a free wave equation.

\subsec{A Lehmann representation for CAdS}

D\"usedau and Freedman \dusedau\ constructed a Lehmann representation
for AdS using the decomposition of the field theory Hilbert space on
$\ads{d+1}$ into representations of $SO(d,2)$. The starting point is
the identity written as a sum over conformal representations.
Labeling the conformal weight by $\Delta$ and the elements of the
representation by $(n,l)$, we have:
\eqn\condecomp{
	{\bbb I} = \int d\Delta \sum_{n,\ell,m}
		\ket{\Delta;n,\ell,m}\bra{\Delta;n,\ell,m}
}
where the energy of $\ket{\Delta;n,\ell,m}$ with respect to global time
is $\omega^\Delta_{n\ell} = \Delta + \ell + 2n$.  In a Lehmann
representation, the two point function of the full (interacting)
scalar fields $\Phi$ can be written in terms of the two-point function
of free scalar fields $\phi$ in representation $\Delta$, integrated
over $\Delta$ with a spectral weight $\rho(\Delta)$:
\eqn\globalordered{
	\langle T\left(\Phi(x)\Phi(y)\right)\rangle
	= \int d\Delta \, \rho(\Delta) \, G_F(\Delta;x-y) \ .
}
Here $G_F$ is the Feynman propagator for a free scalar field
with mass $m^2_\Delta = \Delta(\Delta - d)$.  

Determination of $\rho$ proceeds as in flat space. Consider a 
a canonically normalized free field of conformal weight $\Delta$.  
It follows trivially that 
\eqn\freefcn{\bra{0}\phi(x)\ket{\Delta;n,\ell,m} = \Phi^{+}_{n\ell m}(x)\ }
where $\Phi^{+}$ is the normalizable wavefunction multiplying 
$\hat{a}_{n\ell m}$ in the expansion for $\phi$.
A general interacting scalar field $\Phi$  must have the same matrix
element up to a normalization  factor: 
\eqn\interf{\bra{0}\Phi(x)\ket{\Delta;\ell,n} = 
	N(\Delta) \Phi^{+}_{n\ell}(x)\ ,}
for any $\Delta$.  Eq. \globalordered, with 
$\rho(\Delta) = |N(\Delta)|^2 $,
then follows from inserting \condecomp\ into the right hand side of
the two-point function.

A similar decomposition holds on the boundary, since its Hilbert space
and symmetries are the same.  The coordinate dependence in
\eqn\bdyf{ \bra{0}\CO(b)\ket{\Delta;\ell,n} = \hat{N}(\Delta)
	e^{-i\omega^\Delta_{\ell,n} t} Y_\ell (\Omega)}
is again determined by symmetries.  For a conformal primary operator
$\CO$ with dimension $\Delta(\CO)$, $\hat{N}$ will have support only
at $\Delta = \Delta(\CO)$.  This is compatible with the more
complicated spectral decomposition of the dual operator $\Phi$, so
long as the minimal value of $\Delta$ for which $N(\Delta)$ has
support is $\Delta(\CO)$.  Eq. \boundrestr\ implies that
\eqn\matrixrestr{
	\bra{0}\CO(\Omega,t)\ket{\Delta;\ell,n}
	= \lim_{\rho\to \pi/2} (\cos\rho)^{-\Delta(\CO)}
	\bra{0}\Phi(\rho,\Omega,t)\ket{\Delta;\ell,n}\ .
}
The interacting field $\Phi$ may contain components of many different
conformal weights, but the matrix element
$\bra{0}\Phi(\rho,\Omega,t)\ket{\Delta;\ell,n}$ is a normalizable mode
with fixed conformal weight $\Delta$.  So it falls off at the
boundary as $\cos^\Delta\rho$, and the right side of
\matrixrestr\ vanishes for any $\Delta > \Delta(\CO)$.

Eq.~\matrixrestr\ shows that restricting $\Phi$ to the boundary
isolates the representation of $SO(2,d)$ with the lowest
mass/conformal dimension in the interacting field.  This should be
thought of as a ``single-particle state'' corresponding to $\Phi$,
following the flat space analogy.  The CFT primary $\CO$ carries
precisely this conformal dimension.  The CFT operators function
similarly to the ``in'' and ``out'' states:
\eqn\weakasympt{
	\bra{0}\Phi\ket{\Delta,n,\ell}
	{\buildrel \rho \to \pi/2 \over \longrightarrow}
	(\cos\rho)^{\Delta(\CO)} 
	\bra{0}\CO\ket{\Delta,n,\ell}
}
is the analog of the weak asymptotic condition
\eqn\flatweak{
	\bra{0}\Phi\ket{\Delta,n,\ell}
	{\buildrel t\to-\infty \over \longrightarrow} 
	Z^{-\half}\bra{0}\Phi_{{\rm in}}\ket{\Delta,n,\ell}
}
in flat-space field theory (\cf\ for example \iandz.)

\paragraph{Moral of the story: } The fact that restricting the bulk
interacting fields to the boundary isolates the single particle states
teaches us two very important things.  First we understand why
\bdhmeq\ fails as an attempt to reconstruct local bulk correlators;
the formula does not contain the ``multi-particle part'' 
of the bulk fields.
Secondly, we understand why the S-matrix prescriptions of Sec.~3 and
Sec.~4 work even in the interacting theory; boundary operators really
do create acceptable single-particle asymptotic
states. This feature of the map \oprest\ gives us one way to think about the general problem of reconstructing interacting bulk fields from the boundary.  Presumably, in order to create the ``multi-particle'' pieces of a bulk field we will have to write a CFT operator containing a sum of various products of CFT primaries and descendants.

\subsec{Towards interacting fields}

In this section we will try to construct the interacting field
$\phi$ from the CFT. We will operationally reproduce a spacetime
perturbation expansion via calculations in the large $N$ CFT,
where there is a free-field representation.

\subsubsec{Spectral decomposition and propagators}

Our starting points are the decompositions \freeprim\ and \pfreeprim:
\eqn\gspect{\CO(t,\Omega) = \sum_{n,l} \, \CO(n,\ell; t,\Omega)}
and
\eqn\pspect{\CO(t,\vec{x})= \int dq \, \CO(q; \vec{x},t)\ ,}
which write the CFT operators as sums of appropriate Fourier modes.
$\CO(n,\ell)$ and $\CO(q)$ may be extracted from $\CO$ via a
(nonlocal) transform following from \freeprim\ and \pfreeprim.
In the Poincar\'e case, \pspect\ is an effective decomposition of
$\CO$ onto ``mass-shells'' of fixed $q^2=\omega^2-k^2$.  
Indeed, we can use \pfreeprim\ to suggestively write the two-point
function as:
\eqn\freecfttwo{
\eqalign{
	\langle T\left(\CO(t,\vec{x}) \, \CO(t',\vec{x}')\right)\rangle
	&= \int dq \, 
	\langle T\left(\CO(q;t,\vec{x}) \, \CO(q;t',\vec{x}')\right)\rangle\cr
	&=\frac{1}{2^{2\nu+1}(\Gamma(1+\nu))^2}
		\int dq \, q^{2\nu + 1} \,
		\Delta(q^2;t-t',\vec{x}-\vec{x}')\ 
}}
where $\Delta(q^2)$ is the Feynman propagator for a free
particle with mass $q^2$ on ${\bbb R}^{1,d-1}$.  This
is the spectral representation for the
two-point function of a conformal primary with dimension
$2h_+ = (d/2 + \nu)$ (\cf\ Ref. \capelli).  A similar representation
should exist
for global coordinates.  

The ``transfer matrix'' relation in Sec.~5.1.2 between bulk and
boundary fields was nonlocal in the boundary coordinates.  In
contrast, the large $N$ ``spectral fields'' $\CO(n,\ell)$ and
$\CO(q)$ are related to  bulk free fields by {\it local}
transformations:
\eqn\gspecexp{
\eqalign{
	\Phi(t,\rho,\Omega) &= \Gamma(1+\nu)
	\sum_{n,\ell,m} \left(
	\frac{\Gamma(1+n)}
		{\Gamma(1+n+\nu)}\right)^\half\times\cr
	&\ \ \ \ \ \times
	(\sin\rho)^\ell \, (\cos\rho)^{\Delta} \,
	P_n^{(\ell+\frac{d}{2}-1,\nu)}(\cos 2\rho) \, 
	\CO(n,\ell;t,\Omega)\cr
	& \equiv \sum_{n,\ell} \alpha(n,\ell;\rho) \,  
		\CO(n,\ell;t,\Omega)\ 
}}
and
\eqn\pspecexp{
\eqalign{
	\Phi(z,t,\vec{x}) &= \Gamma(1+\nu)
	\int dq\,  \left( \frac{2}{q}\right)^\nu \, 
		z^{\frac{d}{2}}\,  J_\nu (qz) \,
	\CO(q;t,\vec{x})\cr
	& \equiv \int dq \, \beta(q;z) \, \CO(q;t,\vec{x})\ .
}}
The latter formula implies that free bulk operators arise
from a Bessel transformation of large $N$ spectral fields on the
boundary. 

The simplicity of these relations extends to bulk two-point functions.  If
we attempt to write these using \transf, the time-ordering operator fails
to push through to the boundary fields and a cumbersome expression results.
The spectral fields yield a simpler result:\foot{For simplicity we
focus on the Poincar\'e case.}
\eqn\spectwo{
\eqalign{
	\langle T\left(\Phi(z_1,\b_1)\Phi(z_2,\b_2)\right)\rangle
	&= \int dq_1 \, dq_2 \, \beta(q_1;z_1) \, \beta(q_2;z_2) \,
	\langle T\left(\CO(q_1;\b_1) \, \CO(q_2;\b_2)\right)
		\rangle\cr
	&= {1 \over 2^{2\nu +1} (\Gamma(1 + \nu))^2}
\int dq \, q^{2\nu+1} \, \beta(q;z_1)\, \beta(q;z_2) \,
	\Delta(q^2;\b_1-\b_2)\ . 
}}
Here the only transformation we have done is between bulk radial
coordinate and CFT mass-shell, and the time-ordering remains
untouched.

\subsubsec{Interactions}

Eq.~\oprest\ appears to relate the bulk interacting fields to the
boundary operators.  However, we showed that it does so in a way that
projects onto the non-interacting part of the field.  Is there a
prescription to recover the full bulk fields, or equivalently their
correlators?

There is no reason to expect that the interacting fields can be
linearly expanded in terms of creation and annihilation operators
obeying \freecr\ or \pcr.  Instead, they should contain products of
the free field creation and annihilation operators, corresponding to
``multi-particle'' interacting states.  At least in the semiclassical
(large $N$) limit, this distinction is captured in the relation
between Heisenberg and interaction picture operators:
\eqn\intheis{
	\Phi_H (x,t) = e^{-iH t} e^{i H_0 t} 
		\Phi_I (x,t) e^{-i H_0 t} e^{i H t}\ .
}
Their correlators are related by 
\eqn\ipcorr{
	\langle T\left(\phi_{1,H}(z_1,\b_1)\ldots
		\phi_{n,H}(z_n,\b_n)\right)\rangle
	= \langle T\left(\phi_{1,I}(z_1,\b_1)\ldots
	\phi_{n,I}(z_n,\b_n)
	e^{-i\int dt  {\cal H}_{{\rm int}}(\b,z,t)}\right) \rangle\ 
}
where ${\cal H}_{{\rm int}}$ is the interaction Hamiltonian.  Within
perturbative supergravity, ${\cal H}_{{\rm int}}$ is well-defined and
the interaction terms are suppressed by powers of $1/N$ \Maldconj.

We would like to reproduce this perturbative picture from the boundary
perspective.  In the finite $N$ CFT it is far from clear that there is
a corresponding definition of the ``interaction representation.''
However, just as \ipcorr\ is an expansion around the free limit, we
can attempt a construction about the infinite $N$ CFT. The idea is to
begin with the free-field operators, construct the bulk interaction
picture fields via Eqs. \transf, \gspecexp, or \pspecexp, and then
infer from the bulk lagrangian the corresponding perturbation
expansion in the CFT.

This procedure is most transparent in the language of the spectral
fields.  Suppose that the interaction Hamiltonian is a simple coupling
of three scalars $\phi_i$,\foot{Again, we focus on the Poincar\'e
case.}
\eqn\sampleham{
	H_{{\rm int}} = \frac{\lambda}{N}
	\int dz \, d^{d-1}\vec{x} \, \phi_i(z,\b) \,
	\phi_j(z,\b) \, \phi_k(z,\b)\  .
}
We transfer this to the boundary using the Bessel transformation in
\pspecexp, and integrate over $z$.   This gives:
\eqn\cftsample{H_{{\rm int}}(t)(\{\CO(q)\}) 
		= \int dq_i \, dq_j \, dq_k \, 
        d^{d-1}\vec{x} \, 
	V(q_i,q_j,q_k) \, \CO_i(q_i;t,\vec{x}) \, \CO_j(q_j;t,\vec{x})
	\, \CO_k(q_k;t,\vec{x}) }
where 
\eqn\vdefi{V(q_i,q_j,q_k) = \frac{\lambda}{N}
		\int dz \, \beta(q_i;z) \, \beta(q_j;z) \, 
		\beta(q_k;z)\ }
and $\CO_i(\b)$ are the boundary operators dual to the scalars.
Other terms in the bulk interaction Hamiltonian can
also be converted into functionals of $\CO(q)$ in the
obvious way.  
The bulk correlator \ipcorr\  is then:
\eqn\collcorr{
\eqalign{
	&\langle T\left(\phi_{1,H}(z_1,\b_1)\ldots
		\phi_{n,H}(z_n,\b_n)\right)\rangle = \cr
	&\ \ \ \ \ \int \prod_{i=1}^{n} dq_i \, \beta(q_i;z_i) \,
	\langle T\left( \CO(q_1;\b_1)\ldots\CO(q_n;\b_n)
		e^{-i\int dt H_{{\rm int}} (\{\CO(q)\})}
		\right) \rangle\ .
}}

The expectation value in the second line can be computed
perturbatively in $H_{{\rm int}}$ using Wick's theorem and the
propagator in Eq. \freecfttwo.  This propagator can in turn be derived
by treating $\CO(q)$ as a field with an appropriate lagrangian.
Including interaction terms, we can therefore imagine deriving the
amplitudes perturbatively from
\eqn\cclag{
\eqalign{
	\CL & = \sum_k \int dq_k \, d\b \, 
	\frac{2^{2\nu_k+1}\Gamma(1+\nu_k)^2}{q_k^{2\nu_k+1}}
	\left[ (\p \CO_k(q_k;\b))^2 + q_k^2 \CO_k(q_k;\b)^2 \right]\cr
	&\ \ \ \ \ + 
		\frac{1}{N}\sum_{ijk} 
		\int d\b \, dq_i \, dq_j \, dq_k \,
	V(q_i,q_j,q_k) \, \CO_i(q_i;\b) \, \CO_j(q_j;\b) \,
		\CO_k(q_k;\b) + \ldots\ .
}}
This should be interpreted as a lagrangian for the
``collective fields'' $\CO(q)$, and  
allows us to compute CFT correlators in a power
series in $1/N$.\foot{We are using the term
``collective field theory'' somewhat loosely.
The feature our theory shares with
collective field theory in the usual sense
\refs{\jevicki} is that we take an overcomplete
basis of operators and promote them to
independent lagrangian fields.  And,
as in \refs{\jevicki}, this lagrangian
gives a $1/N$ expansion for exact correlators
of the theory.  It is clearly desirable to
relate these two theories.}

The dimension of $\CO$ is encoded
in the normalization factor
\eqn\knorm{
	A(q,\nu) = \frac{2^{2\nu+1}\Gamma(1+\nu)^2}{q^{2\nu+1}}\ ,
}
and the operator product coefficients for the
primary operators are encoded in the three-point
couplings $V(q_1,q_2,q_3)$.  

\subsec{Interpretation}

Our collective field lagrangian \cclag\ is local in spacetime after
transforming $q$ to $z$ via the kernel $\beta$, and completely
captures the leading perturbative supergravity.  It is true that we
have constructed this theory from supergravity by fiat, but we have
thus found a fairly simple but not {\it a priori} obvious integral
transformation between the resulting collective fields and the
$N=\infty$ CFT variables.  We can now pose sharp questions directly
within the CFT about the reconstruction of local spacetime physics.
Can we reproduce a collective field lagrangian like \cclag\ (and
therefore bulk supergravity) directly from the gauge theory via
standard large $N$ techniques?  Can such a representation of the
theory be found at finite $N$?  In particular, the loop equations for
4d $N=4$ Yang-Mills theory, which we might hope reproduces string
dynamics on $\ads{5}\times S^5$, should contain our collective field
theory in some $\apr\to 0$ approximation.  Furthermore, the integral
transformation between bulk and boundary fields \pspecexp\ is strongly
reminiscent of the relation between correlators of collective fields
and tachyon fields in the $c=1$ matrix model of 2D string theory
\gregandnati.  We suspect that this similarity is not
accidental.\foot{In fact the authors of \oferetal\ have suggested
that the $c=1$ matrix model is a holographic description of
1+1-dimensional string theory, precisely in the fashion of the AdS/CFT
correspondence.}

It would be interesting to see how the discussions in
\refs{\BDHM,\BKLT} relate to this manifestly local description.  In
those papers, it was found that for classical probes, bulk locality in
$z$ translated to a sort of ``scale-size locality'' in the large-N
boundary theory.  In Eq. \pspecexp, $q$ functions as a scale,
selecting a mass shell in the Fourier expansion of the primary
operator $\CO$.  This is suggestive, but does not give a precise
translation of bulk radial locality into CFT scales because because
$q$ is related to $z$ by an integral transform.

It is also amusing to note that the steps in section 5.3 can be reproduced
in other spaces besides AdS, and used to define a ``boundary theory" from
a given bulk theory.  One simple example  defines a boundary lagrangian on
the ${\bbb R}^{d-1}$ boundary of the upper half space of ${\bbb R}^d$.
This observation may be relevant to exploration of holography in more
general contexts.

Of course \cclag\ cannot be the whole story.  The expansion it
provides in $1/N$ and $\gym^2$ will undoubtedly be asymptotic at best,
and the Hilbert space of this collective field theory is not the same
as the Hilbert space of the finite-$N$ CFT.\foot{As an example, in 
$\ads{3}$, $N$ is related to the central charge of the theory.
We are then trying to capture the dynamics of a theory with finite $c$
via an infinite-$c$ ``free field'' description.}  However, this
statement is precisely what makes the collective field theory \cclag\
an interesting starting point for studying bulk dynamics.  Classical
gravity and its na\"\i ve semiclassical extension are afflicted with
well-known pathologies.  The degree to which our collective field
theory is a poor approximation to the exact CFT should be precisely
the degree to which our assumptions about classical and semiclassical
gravity are invalid; the full CFT, which is manifestly well-defined,
will remove the pathologies.  How it does so can in principle be
learned by understanding the difference between our collective field
description and the full conformal field theory.  This difference
will contain the non-locality resulting from the
holographic description.

\newsec{Conclusion}

In this article we have interpreted the correlation functions of the
CFT duals of AdS spaces in terms of bulk S-matrices and transition 
amplitudes.  First, vacuum
correlators of the CFT were expressed in the bulk as truncated n-point
functions convolved against non-normalizable on-shell modes.  We
interpreted these expressions as an S-matrix for AdS arising from a limit
of scattering
in asymptotically flat space from an AdS bubble.  In the free limit,
fields in AdS spacetime possess a class of normalizable, fluctuating
solutions.  A traditional LSZ prescription would compute transition
amplitudes between these states.  We showed that the usual LSZ
framework fails in global AdS spacetime, essentially because the
periodicity of geodesics obscures the definition of asymptotic states.
We then avoided this subtlety by partitioning AdS into patches and
showed that CFT correlation functions compute transition amplitudes
between suitable states defined on these patches.  These states
correspond to boundary conditions on the early and late time surfaces
of a patch and, unusually, were created and annihilated by the action
of operators on the vacua of earlier and later patches.

Finally, we have tried to directly reconstruct local bulk operators
from the boundary CFT.  In the $N\to\infty$ limit, this reconstruction
is straightforward.  But at finite $N$, single CFT primaries capture
only the ``one-particle'' parts of the dual bulk operators, making
reconstruction a nontrivial, nonlinear process.  We proposed a
``collective field theory'' which reproduces CFT correlators order by
order in $\gym$ and $1/N$, and manifestly reproduces local,
perturbative supergravity after a straightforward integral
transformation.  However, this theory is not the full CFT and we
expect that the perturbation series is asymptotic at best.

It is worth noting that many features of the collective field
description espoused in Sec.~5 break a striking resemblance to the
matrix models of 2d string theory.  This similarity has been an
important source of inspiration in the development of the AdS/CFT
correspondence \refs{\gkp,\emilmat,\BKL} and we see it as an
interesting guide for future work.

We have argued that it is possible to construct S-matrices and
transition amplitudes for spacetime states from the conformal field
theory dual.  We expect that these amplitudes cannot at the
fundamental level be obtained from a local bulk quantum gravity
theory, and it would be interesting to learn how the violation of
locality is manifested.  One approach is to look for breakdown of
spacetime locality in deviations of the finite $N$ dynamics from the
local-by-construction collective field description proposed in this
article.  This field theory constructs CFT correlators in a
power-series expansion in $1/N$ and $\gym^2$.  Perhaps analyticity of
this power series is in itself related to locality.  It is worth
recalling the usual story where ``mean field'' assumptions can be
destroyed by fluctuations. Perhaps locality and causality are
``mean-field'' properties.  This idea is in line with the suggestions
in Refs.~\refs{\jacobson,\BDHM,\BKLT,\emilstat} that local spacetime
and causal properties are somehow thermodynamic in character.

\bigskip\bigskip\centerline{{\bf Acknowledgments}}\nobreak

We have enjoyed extensive conversations and collaboration with Per
Kraus on some of the material presented here.  We thank O. Aharony,
T. Banks, S. Das, R. Gopakumar, A. Grant, D. Gross, E. Martinec,
S. Mathur, J. Polchinski, A. Rajaraman, S.-J. Rey, S. Ross, and
E. Silverstein for useful conversations that have influenced many
ideas in this work.  V.B. is supported by the Harvard Society of
Fellows and by NSF grants NSF-PHY-9802709 and NSF-PHY-9407194.
S.B.G. is supported in part by DOE contract DE-FG-03-91ER40618.
A.L. is supported by NSF grant NSF-PHY-9802709.  V.B. is grateful to
Rutgers University and Seoul National University for hospitality while
some of this work was being carried out.  A.L. is grateful to the
Depts. of Physics at U.C. Santa Barbara and U.C. Berkeley, the theory
groups at SLAC and LBNL, and the Center for Geometry and Theoretical
Physics at Duke University for their hospitality at various times
during the course of this project.  This effort was initiated at the
1998 summer workshop on string theory at the Aspen Center for Physics.

\listrefs
\end